\documentclass{aa}  

\usepackage{graphicx}
\usepackage{subcaption}
\usepackage{amsmath}
\usepackage{txfonts}
\usepackage{hyperref}
\usepackage{subcaption}

\usepackage{suffix}
\usepackage[printonlyused]{acro}

\acsetup{
  single        = false,
  list/sort     = true,
  cite/group    = true,
  cite/group/cmd = \citealp,
  cite/group/pre = {; },
}

\DeclareAcronym{EVE}{
  short = EVE,
  long = \textit{Extreme Ultraviolet Variability Experiment},
  cite = {Woods:2012}}

\DeclareAcronym{AIA}{
  short = AIA,
  long = \textit{Atmospheric Imaging Assembly},
  cite = {Lemen:2012}}

\DeclareAcronym{HMI}{
  short = HMI,
  long = \textit{Helioseismic Magnetic Imager},
  cite = {Scherrer:2012}}
  
\DeclareAcronym{SDO}{
  short = SDO,
  long = \textit{Solar Dynamics Observatory},
  cite = {Pesnell:2012}}

\DeclareAcronym{STEREO}{
  short = STEREO,
  long = \textit{Solar Terrestrial Relations Observatory},
  cite = {Kaiser:2008}}

\DeclareAcronym{SECCHI}{
  short = SECCHI,
  long = \textit{Sun Earth Connection Coronal and Heliospheric Investigation},
  cite = {Howard:2008}}
  
\DeclareAcronym{SCIP}{
  short = SCIP,
  long = Sun Centered Imaging Package}
  
\DeclareAcronym{MGN}{
  short = MGN,
  long = multi-scale Gaussian normalisation,
  cite = {Morgan:2014}}

\DeclareAcronym{EUVI}{
  short = EUVI,
  long = EUVI}

\DeclareAcronym{CME}{
  short = CME,
  short-plural-form = CMEs,
  long = coronal mass ejection,
  long-plural-form = coronal mass ejections}
  
\DeclareAcronym{PIL}{
  short = PIL,
  short-plural-form = PILs,
  long = polarity inversion line,
  long-plural-form = polarity inversion lines}
  
\DeclareAcronym{LOS}{
  short = LOS,
  short-plural-form = LOS,
  long = line of sight,
  long-plural-form = lines of sight}
  
\DeclareAcronym{FOV}{
  short = FOV,
  short-plural-form = FOV,
  long = field of view,
  long-plural-form = fields of view}
  
\DeclareAcronym{UV}{
  short = UV,
  short-plural-form = UV,
  long = ultraviolet,
  long-plural-form = ultraviolet}

\DeclareAcronym{EUV}{
  short = EUV,
  short-plural-form = EUV,
  long = extreme ultraviolet,
  long-plural-form = extreme ultraviolet}
  
\DeclareAcronym{GONG}{
  short = GONG,
  long = \textit{Global Oscillations Network Group}}
  
\DeclareAcronym{DST}{
  short = DST,
  long = \textit{Dunn Solar Telescope}}

\DeclareAcronym{IBIS}{
  short = IBIS,
  long = \textit{Interferometric Bidimensional Spectropolarimeter},
  cite = {Cavallini:2006,Reardon:2008}}

\DeclareAcronym{FIRS}{
  short = FIRS,
  long = \textit{Facility Infrared Spectropolarimeter},
  cite = {Jaeggli:2011}}
  
\DeclareAcronym{EIT}{
  short = EIT,
  long = \textit{Extreme Ultraviolet Imaging Telescope},
  cite = {Delaboudiniere:1995}}
 
\DeclareAcronym{ROSA}{
  short = ROSA,
  long = \textit{Rapid Oscillations in the Solar Atmosphere},
  cite = {Jess:2010}}
  
\DeclareAcronym{SPINOR}{
  short = SPINOR,
  long = \textit{Spectro-Polarimeter for Infrared and Optical Regions},
  cite = {Socas:2006}}  
 
\DeclareAcronym{SOHO}{
  short = SOHO,
  long = \textit{Solar and Heliospheric Observatory},
  cite = {Domingo:1995}}
 
\DeclareAcronym{MDI}{
  short = MDI,
  long = \textit{Michelson Doppler Imager},
  cite = {Scherrer:1995}}

\DeclareAcronym{NMSU}{
  short = NMSU,
  long = New Mexico State University}

\DeclareAcronym{NSO}{
  short = NSO,
  long = \textit{National Solar Observatory}}
  
\DeclareAcronym{TAC}{
  short = TAC,
  long = time allocation committee}
  
\DeclareAcronym{DEM}{
  short = DEM,
  short-plural-form = DEMs,
  long = Differential Emission Measure,
  long-plural-form = Differential Emission Measures}
  
\DeclareAcronym{PI}{
  short = PI,
  short-plural-form = PI,
  long = Principle Investigator,
  long-plural-form = Principle Investigators}  
  
\DeclareAcronym{IDL}{
  short = IDL,
  long = Interactive Data Language}
  
\DeclareAcronym{bb}{
  short = bb,
  long = broadband}  
  
\DeclareAcronym{nb}{
  short = nb,
  long = narrowband}    
  
\DeclareAcronym{IPM}{
  short = IPM,
  long = interplanetary medium}  
  
\DeclareAcronym{AO}{
  short = AO,
  long = adaptive optics,
  cite = {Rimmele:2004}} 
  
\DeclareAcronym{IR}{
  short = IR,
  long = infrared}  

\DeclareAcronym{halpha}{
  short = H-$\alpha$,
  long = Hydrogen-$\alpha$}

\DeclareAcronym{FPI}{
  short = FPI,
  long = Fabry-P\'erot}
  
\DeclareAcronym{DWDM}{
  short = DWDM,
  long = dense wavelength division multiplexing}  

\DeclareAcronym{CCD}{
  short = CCD,
  long = charge-coupled device}   
  
\DeclareAcronym{HI}{
  short = HI,
  long = Heliospheric Investigation}   
  
\DeclareAcronym{ROI}{
  short = ROI,
  long = region-of-interest}    
  
\DeclareAcronym{POV}{
  short = POV,
  long = point-of-view}  
  
\DeclareAcronym{MHS}{
  short = MHS,
  long = magnetohydrostatic}
  
\DeclareAcronym{MHD}{
  short = MHD,
  long = magnetohydrodynamic}  
  
\DeclareAcronym{RMHD}{
  short = RMHD,
  long = radiative magnetohydrodynamic}  
  
\DeclareAcronym{DOT}{
  short = DOT,
  long = \textit{Dutch Open Telescope},
  cite = {Rutten:1997}}  
  
\DeclareAcronym{BBSO}{
  short = BBSO,
  long = \textit{Big Bear Solar Observatory}}
  
\DeclareAcronym{NST}{
  short = NST,
  long = \textit{New Solar Telescope},
  cite = {Goode:2012}}

\DeclareAcronym{SOT}{
  short = SOT,
  long = \textit{Solar Optical Telescope},
  cite = {Tsuneta:2008}}
  
\DeclareAcronym{hinode}{
  short = Hinode,
  long = Hinode,
  cite = {Kosugi:2007}}
  
\DeclareAcronym{GST}{
  short = GST,
  long = \textit{Goode Solar Telescope}}  
  
\DeclareAcronym{SST}{
  short = SST,
  long = \textit{Swedish 1-m Solar Telescope},
  cite = {Scharmer:2002}}  
  
\DeclareAcronym{TRACE}{
  short = TRACE,
  long = \textit{Transition Region and Coronal Explorer},
  cite = {Handy:1999}}  
  
\DeclareAcronym{PCTR}{
  short = PCTR,
  long = \textit{prominence-corona-transition-region}}
  
\DeclareAcronym{DKIST}{
  short = DKIST,
  long = Daniel K. Inouye Solar Telescope}
  
\DeclareAcronym{RTI}{
  short = RTI,
  long = Rayleigh-Taylor instability}
  
\DeclareAcronym{SSW}{
  short = SSW,
  long = SolarSoftWare,
  cite = {Freeland:1998}}    
  
\DeclareAcronym{LTE}{
  short = LTE,
  long = local thermodynamic equilibrium}
  
\DeclareAcronym{NLTE}{
  short = NLTE,
  long = non-"local thermodynamic equilibrium"}
  
\DeclareAcronym{FTS}{
  short = FTS,
  long = Fourier Transform Spectrometer,
  cite = {Kurucz:1984}}
  
\DeclareAcronym{RTE}{
  short = RTE,
  long = radiative transfer equation}
  
\DeclareAcronym{CRD}{
  short = CRD,
  long = complete frequency redistribution}
  
\DeclareAcronym{PRD}{
  short = PRD,
  long = partial frequency redistribution}
  
\DeclareAcronym{BCM}{
	short = BCM,
	long = Beckers' cloud model,
	cite={Beckers:1964}}
	
\DeclareAcronym{HSRA}{
	short = HSRA,
	long = Harvard Smithsonian Reference Atmosphere,
	cite={Gingerich:1971}}
	
\DeclareAcronym{NICOLE}{
	short = NICOLE,
	long = Non-LTE Inversion COde using the Lorien Engine,
	cite={Socasnavarro:2015}}
	
\DeclareAcronym{SIR}{
	short = SIR,
	long = Stokes Inversion based on Response functions,
	cite={Ruizcobo:1992}}
	
\DeclareAcronym{HAZEL}{
	short = HAZEL,
	long = Hanle and Zeeman Light,
	cite={AsensioRamos:2008}}
	
\DeclareAcronym{BPSS}{
	short = BPSS,
	long = bald patch separatrix surface,
	cite={Bungey:1996}}
	
\DeclareAcronym{EIS}{
	short = EIS,
	long = \textit{EUV Imaging Spectrometer},
	cite={Culhane:2007}}

\DeclareAcronym{FWHM}{
  short = FWHM,
  long = full width at half maximum}

\DeclareAcronym{AMRVAC}{
	short = MPI-AMRVAC,
	long = \textit{Adaptive Mesh Refinement Versatile Advection Code},
	cite = {Keppens:2012,Porth:2014,Xia:2018,Keppens:2020}}

\DeclareAcronym{AMR}{
	short = AMR,
	long = adaptive mesh refinement}

\DeclareAcronym{CCI}{
	short = CCI,
	long = Convective Continuum Instability}

\DeclareAcronym{BV}{
	short = BV,
	long = Brunt-V\"ais\"al\"a}

\DeclareAcronym{TVDLF}{
	short = TVDLF,
	long = Total Variation Diminishing Lax-Friedrich}

\DeclareAcronym{TI}{
	short = TI,
	long = Thermal Instability}

\DeclareAcronym{TNE}{
	short = TNE,
	long = thermal non-equilibrium}
	
\DeclareAcronym{MALI}{
    short = MALI,
    long = Multilevel Accelerated Lambda Iteration}
    
\DeclareAcronym{yt}{
    short = yt,
    long = yt-project,
    cite = {Turk:2011}}
    
\DeclareAcronym{GL}{
    short = GL,
    long = Gauss-Legendre}
    
\DeclareAcronym{CM}{
    short = CM,
    long = classically mounted}
    
\DeclareAcronym{MULTI3D}{
    short = MULTI3D,
    long = \textit{multi-level non-LTE 3D},
    cite = {Leenaarts:2009}}

\DeclareAcronym{EB}{
    short = EB,
    long = \textit{Eddington-Barbier}}
    
\DeclareAcronym{KDE}{
    short = KDE,
    long = kernel density estimate}
    
\DeclareAcronym{NLFFF}{
    short = NLFFF,
    long = nonlinear force-free field}

\usepackage[normalem]{ulem}
\usepackage[x11names]{xcolor}
\usepackage{academicons}
\usepackage[flushleft]{threeparttable}
\usepackage{booktabs,caption}
\definecolor{orcidlogocol}{HTML}{A6CE39}

\newcommand{\orcid}[1]{\href{https://orcid.org/#1}{\textcolor[HTML]{A6CE39}{\aiOrcid}}}

\usepackage{xcolor}
\usepackage{soul}
\definecolor{tumbleweed}{rgb}{0.87, 0.67, 0.53}

\begin{document}

   \title{Merging plasmoids and nanojet-like ejections in a coronal current sheet}

   \author{Samrat Sen$^*$ \inst{1, 2} \href{https://orcid.org/0000-0003-1546-381X}{\includegraphics[scale=0.05]{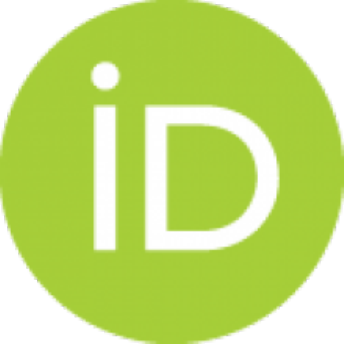}}, F. Moreno-Insertis$^{**}$ \inst{1, 2},
          }

   \institute{
   Instituto de Astrof\'{i}sica de Canarias, 38205 La Laguna, Tenerife, Spain
   \and
   Universidad de La Laguna, 38206 La Laguna, Tenerife, Spain 
   \\
   $^*$\email{samrat.sen@iac.es; samratseniitmadras@gmail.com}
   \\
   $^{**}$\email{fmi@iac.es}
                      }

   \date{Received: XXXX; accepted: XXXX}
   \date{}
 
  \abstract
   {Forced magnetic reconnection is triggered by external perturbations, which are ubiquitous in the solar corona. This process plays a crucial role in the energy release during solar transient events, which are often associated with electric current sheets (CSs). The CSs can often disintegrate through the development of the tearing instability, which may lead to the formation of plasmoids (magnetic islands, or twisted flux ropes in 3D) in the non-linear phase of evolution. However, the complexity of the dynamics, and the magnetic and thermodynamic evolution due to the coalescence of the plasmoids are not fully understood.}
   {We aim to explore the departure of the CS from its equilibrium configuration 
 through a reconnection process characterized by the formation of plasmoids with a guide field, their mutual approach and eventual coalescence, and to investigate the complex magnetic topology and thermodynamic evolution in and around the merged plasmoids.}
   {We used a resistive magnetohydrodynamic simulation of a 2.5D current layer embedded in a stratified medium in the solar corona, incorporating field-aligned thermal conduction using the open-source code MPI-AMRVAC. Multiple levels of adaptive mesh-refined grids are used to resolve the fine structures that result during the evolution of the system.}   
   {The instability in the CS is triggered by imposing impulsive velocity perturbations concentrated at three different locations in the upper half along the CS plane; this leads to the formation of plasmoids and their later coalescence. We demonstrate that a transition from purely 2D reconnection to 2D reconnection with guide field takes place at the interface between the plasmoids as the latter evolve from the pre-merger to the merged state. The small-scale, short-lived, and collimated outflows during the merging process share various physical properties with the recently discovered nanojets. The subsequent thermodynamic change within and outside the merged plasmoid region is governed by the combined effect of Ohmic heating, thermal conduction, and expansion/contraction of the plasma.} 
   {Our results imply that impulsive perturbations in coronal CSs can be the triggering agents for plasmoid coalescence, which leads to the subsequent magnetic, and thermodynamic change in and around the CS.}
 

\titlerunning{Merging plasmoids in a coronal current sheet}
\authorrunning{Sen \& Moreno-Insertis}

\maketitle

\section{Introduction} 

Magnetic reconnection is a fundamental process in which the change of magnetic topology is associated with the conversion of magnetic energy into thermal and kinetic energy, and possibly acceleration of charged particles. This is believed to play an important role in different types of eruptive phenomena in the solar atmosphere, e.g., flares \citep{1939ApJ....89..555G, 1947MNRAS.107..338G, 1948MNRAS.108..163G, 2000mare.book.....P, 2020JGRA..12525935H}, coronal-mass-ejections \citep{1995GeoRL..22.1753G, 2003JGRA..108.1023S, 2012ApJ...760...81K}, coronal jets 
\citep{Yokoyama_Shibata:1996, Pariat_etal:2009, Moreno-Insertis_Galsgaard:2013, Archontis_Hood_2013, Wyper_DeVore:2016}. These phenomena are often associated with the electric current sheets (CSs) that are formed when opposite magnetic field lines come in close proximity of each other and high current density regions are formed and confined in a surface. 

It was first suggested by \cite{1963PhFl....6..459F} that reconnection may be triggered through small perturbations of a current layer, leading to fragmentation of the CS as part of the phenomenon known as tearing instability. Using linear analysis in 2D, \cite{2007PhPl...14j0703L} showed that a chain of magnetic islands could be formed in a single CS due to the development of the tearing instability. These magnetic islands, also called plasmoids, may gradually grow in size in the nonlinear regime under favourable conditions; they are basically a bundle of magnetic field lines winding about a common center that enclose plasma within them, and magnetically isolated from their surroundings. In 3D, the plasmoids adopt the form of twisted magnetic flux ropes. \cite{Bhattacharjee2009, Huang2010, Shen2011, Barta2011, Mei2012} advanced the analysis of the TI, investigating the plasmoid instability in CSs which are not subject to slow Sweet-Parker reconnection, but are inherently unstable to the formation of plasmoids for high Lundquist number ($\gtrapprox10^4$). Through 2D numerical simulations, they showed that a single reconnecting CS can break up into multiple interacting reconnection sites. The effect of temperature-dependent resistivity and thermal conduction for the fast reconnection process in a single CS has been studied by \cite{Ni2012} in a 2D geometry; they found that the temperature enhancement is larger inside the magnetic islands than at the X-points. The inclusion of nonadiabatic effects like radiative cooling and background heating can lead to a thermal runaway process as reported by \cite{Sen2022}, in which cool condensations are formed in a coupled tearing-thermal unstable process and are trapped inside the plasmoids. The islands formed in the CS by the initial reconnection event are attracted to each other due to the parallel currents at the O-points in their center. This results in the coalescence of the islands in a further reconnection event, as also proposed by \cite{Schumacher1997}. The self-similar structure of the CS connecting coalescing islands is the mechanism by which the plasmoid instability accelerates reconnection to a fast regime \citep{Bhattacharjee2009, Huang2010}, and may contribute significantly to the energy release. \cite{Sen2023} extended the study of  coupled tearing-thermal evolution in 3D geometry incorporating the non-adiabatic effects of radiative cooling, background heat, and thermal conduction. This showed the development of magnetic flux ropes, and the formation of cool-condensed plasma in the vicinity of the fragmented CS due to reconnection-driven thermal instability. CSs in the solar corona can be perturbed by external MHD flows, like linear waves or shock waves, leading to forced magnetic reconnection. Observations of these types of events are reported in \citet{Farnik1983, Wang2001, Ofman2018, Zhou2020}, and references therein, for the flare scenario; however, they have received limited attention in numerical models to date. Examples include \cite{SakaiWashimi1982, Sakai1983, Odstrcil1997, Potter2019}, all of which used (localized) anomalous resistivity to facilitate the reconnection, and \cite{Mondal2024} who had uniform resistivity in their model. However, none of the above numerical models focuses on the magnetic and thermodynamic evolution due to merging of plasmoids. This is an important aspect in the plasmoid-mediated fast reconnection scenario, and therefore it warrants deeper investigation. 

In the present paper we investigate the process of merging of two plasmoids created by tearing of an initially straight CS. To that end, we use a bi-directional velocity pinch concentrated at different locations in the CS plane so that two plasmoids are created that subsequently suffer a process of coalescence. This is different to the case of a uni-directional velocity pulse applied at the CS center as prescribed by \cite{Mondal2024}. The use of an external velocity pulse in our model is more realistic for transient disturbances than a periodic function in time as prescribed by \cite{Potter2019}. Additionally, we also incorporate a stratified atmosphere with solar gravity and thermal conduction in our model. The paper is organized as follows. In Section \ref{sec:setup}, we describe the numerical model with the initial configuration and the algorithmic aspects of the boundary conditions. In Section \ref{sec:results}, the main results of the study and their analysis are reported. Section \ref{sec:summary} addresses the significance of the work for a typical coronal medium, summarizes the key findings, the scope of future improvement, and finally concludes how the findings may be useful for future studies.  
\begin{figure*}
    \centering
    \includegraphics[width=0.7\textwidth]{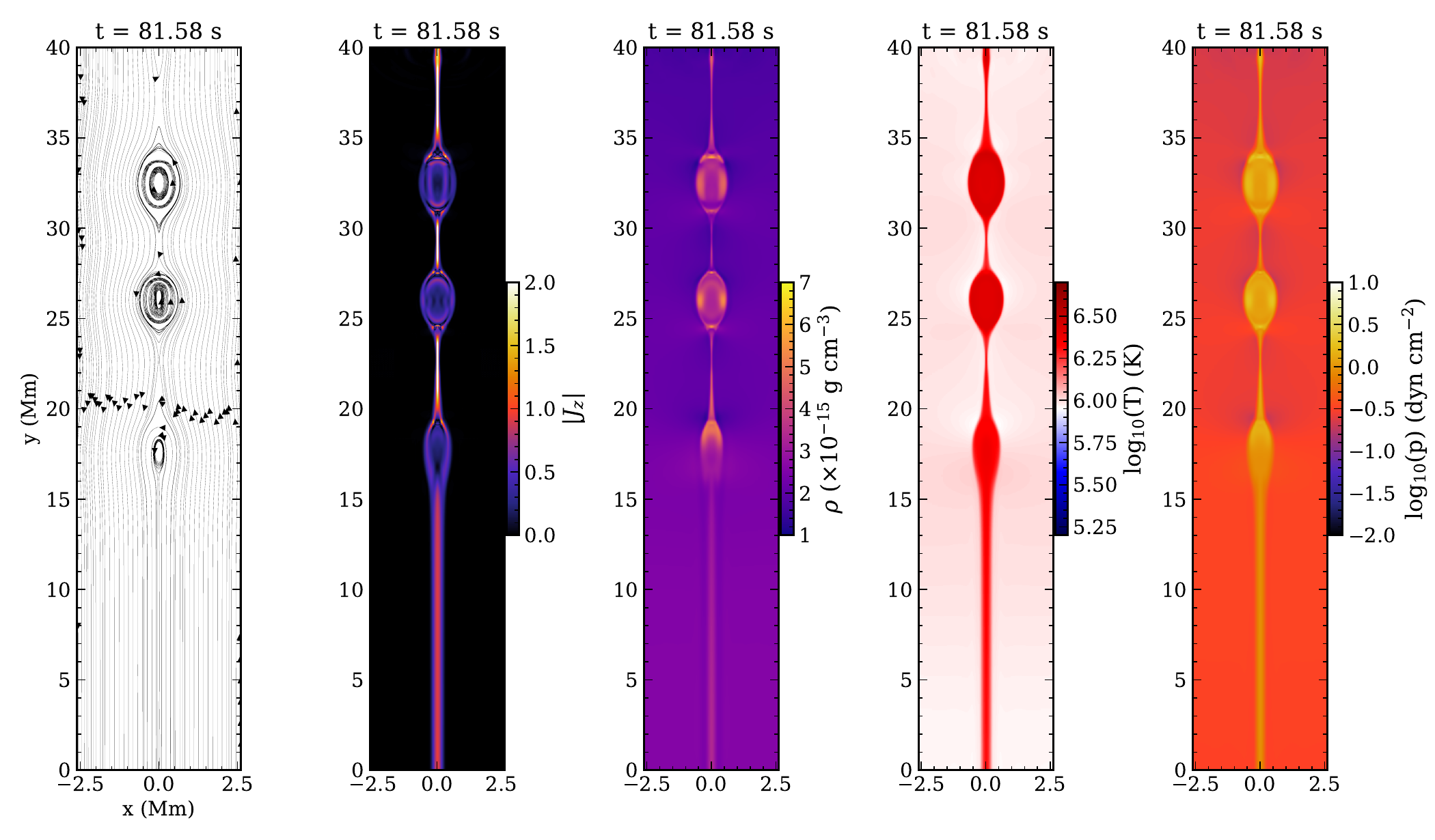}
    \includegraphics[width=0.7\textwidth]{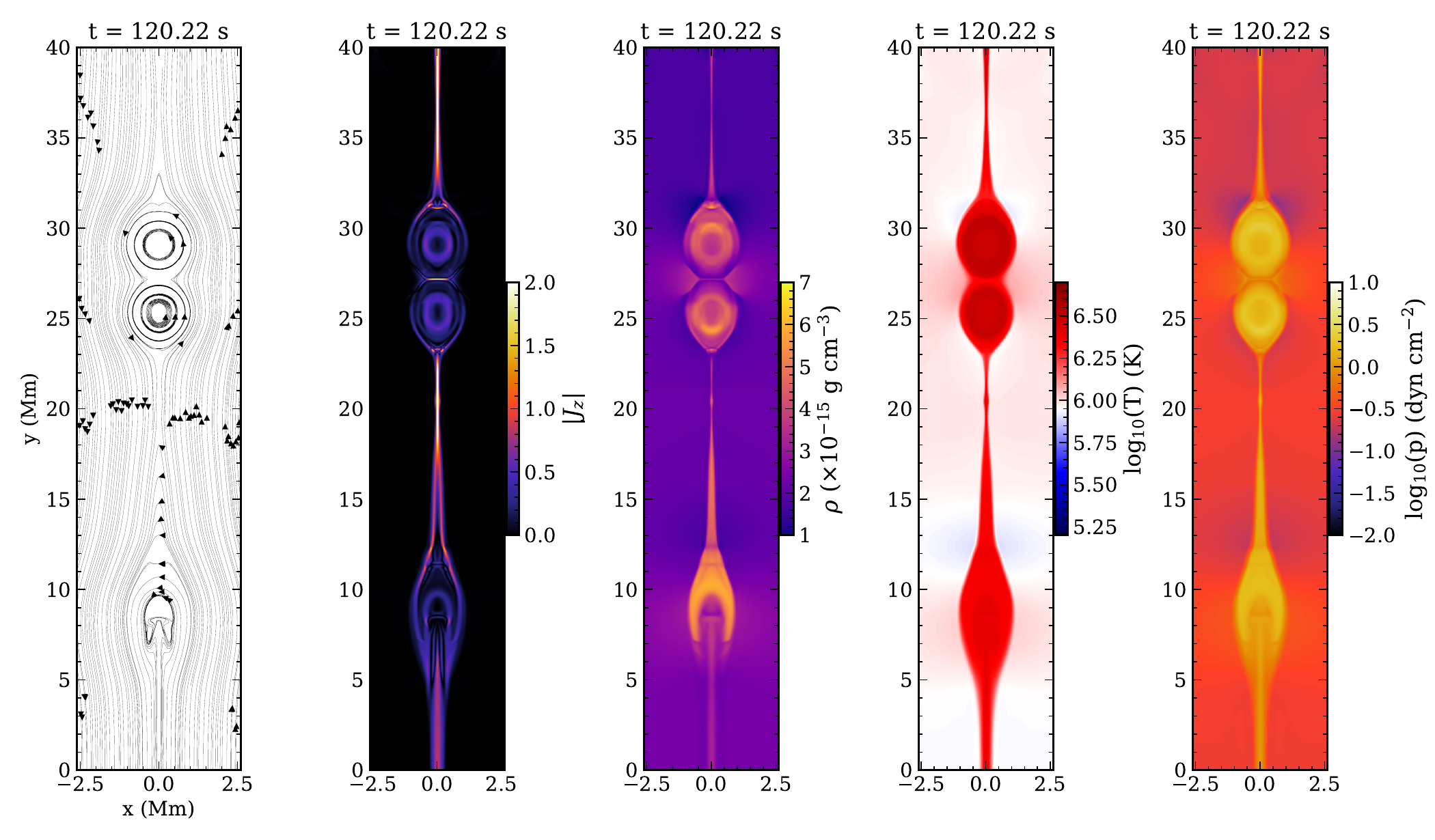}
    \includegraphics[width=0.7\textwidth]{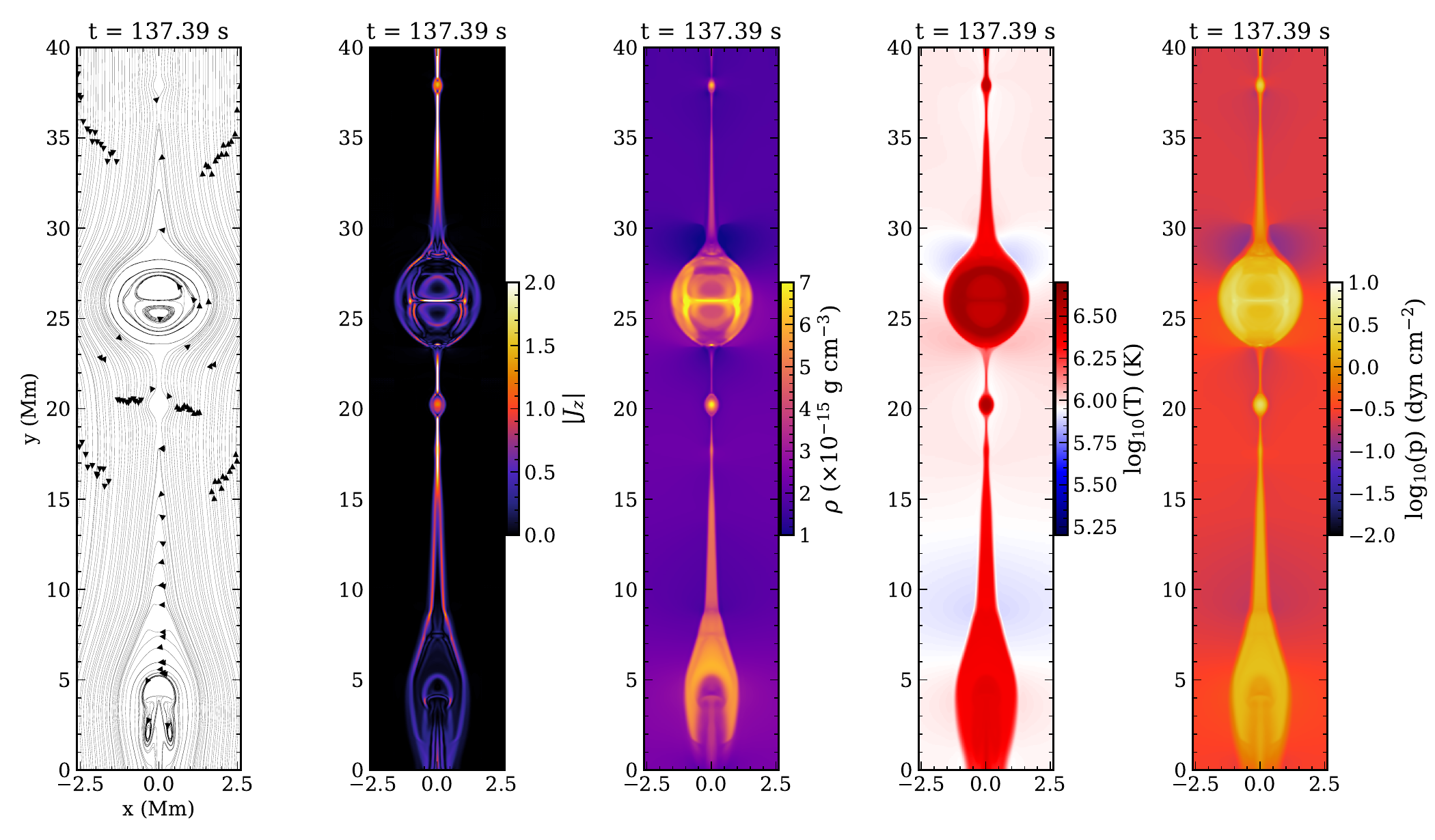}
    \caption{Spatial variations of magnetic field lines projected at the $x-y$ plane, absolute current density ($|J_z|$),  plasma density, temperature, and thermal pressure are shown from left to right at each panels respectively. The evolution of the above quantities for three different times, $t=81.58$, $120.22$, and $137.39$ s are shown at the top, middle and bottom panels respectively. The units of $|J_z|$ is chosen arbitrary and the maximum saturation limit of the color bar is chosen as $2$ for a better visualization of the enhanced current structures. The associated animation (`merging\_plasmoids.mp4') corresponding to this figure is available online.}
    \label{fig:combined}
\end{figure*}

\begin{figure}
    \centering
    \includegraphics[width=1\columnwidth]{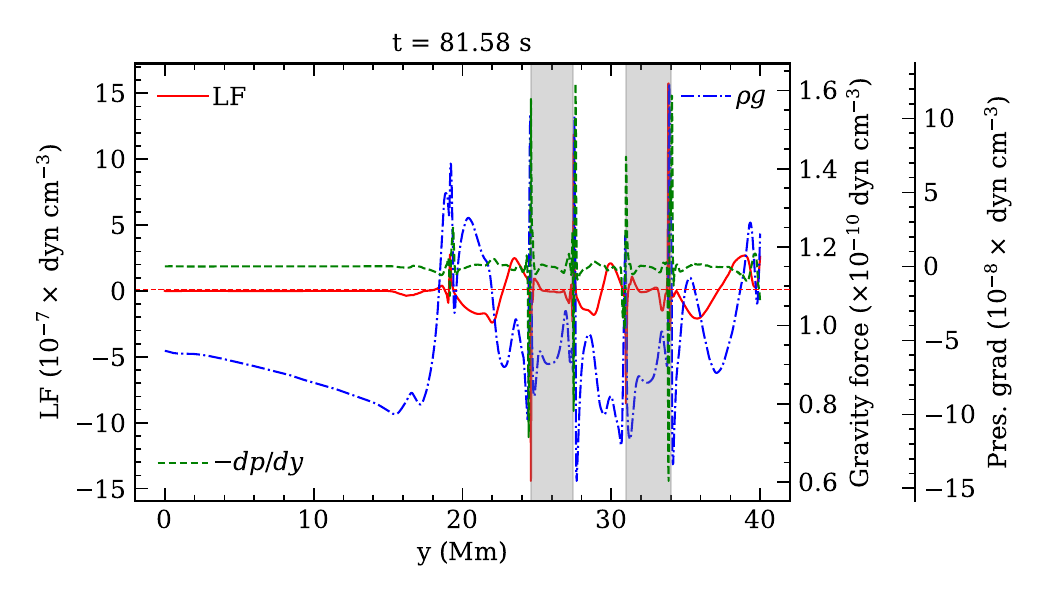}
    \caption{Distribution of the vertical components of Lorentz force (LF), pressure gradient, and gravity force along $y$ at $x=0$ for $t=81.58$ s. The regions shaded in gray cover the vertical extent of the lower and upper plasmoids at $x=0$. The ordinate scale is different for each curve, as shown in the figure. The horizontal red dashed line is the marker of the zero LF.}
    \label{fig:forces}
\end{figure}

\section{The physical and numerical framework of the model}
\label{sec:setup}

To investigate the small-scale dynamics, and the magnetic and thermodynamic evolution in a plasmoid-producing coronal CS, we set up a resistive MHD simulation in a 2.5D cartesian geometry using the MPI-Adaptive Mesh Refinement Versatile Advection Code (MPI-AMRVAC) \citep{2014ApJS..214....4P, 2018ApJS..234...30X, keppens2021, keppens2023}. 
We assume the initial distribution and evolution of all quantities to be
invariant in the $z$-direction, but the magnetic field and the velocity
vectors can have non-zero $z-$components. The initial magnetic field configuration depends on the horizontal coordinate $x$ only:
\begin{align} \label{eq:B_x}
    B_x &= 0,\\ \label{eq:B_y}
    B_y &= B_0\ \tanh\left(\frac{\displaystyle x}{\displaystyle\lambda}\right),\\ \label{eq:B_z}
    B_z &= B_0\ \text{sech}\left(\frac{\displaystyle x}{\displaystyle\lambda}\right),
\end{align}
which corresponds to a current sheet (CS) that extends all along the $y-$direction, and covers a distance of order $\lambda$ along the $x$-axis. The module of the magnetic field is $|B_0|$ in the whole domain and its direction rotates in the vertical $y-z$ plane from $180$ to $0$~degrees with respect to the positive $y$~axis as one crosses the current sheet from left to right along the $x$~axis. The spatial domain of the simulation spans between $x=-20$ to $20$~Mm, and $y=0$ to 40~Mm. We use a typical coronal magnetic field strength, $B_0=6$~G, and $\lambda = 100$~km. The initial pressure and density distributions are taken to correspond to a hydrostatic stratification of uniform temperature $T_0$: 
\begin{align}
       p(y,t=0) = p_0\ \exp\left(-\frac{y}{H}\right)\,,\\
\noalign{\vskip 2mm}
    \rho(y,t=0) = \rho_0\ \exp\left(-\frac{y}{H}\right)\,,
\end{align}
with $H$, the scale-height, given by $H= {\cal R}\,T_0/(g\,\mu)$,  $g$ being the solar gravity and ${\cal R}$ the gas constant; we will use standard coronal values for the temperature, $T_0 = 1$~MK, and for the atomic mass per particle, $\mu=0.6$~g~mol$^{-1}$, which yield $H\approx 50$~Mm. The pressure and density at the basis ($y=0$) are:
$p_0 = 0.42$~dyne~cm$^{-2}$, and $\rho_0=3.14 \times 10^{-15}$~g~cm$^{-3}$. The plasma-$\beta$ varies between $0.30$ and $0.18$ from $y=0$ to $y=40$~Mm. The base resolution of the simulation is taken as $384$~grid cells along both the $x$ and $y$ directions, with, additionally, two levels of AMR, so  that the smallest cell size becomes $\approx 26.04$~km in either direction. The (de-)refinement is based on the errors estimated by the gradients of the instantaneous density and magnetic field components at each time step \citep{1987:Lohner}. We use a uniform diffusivity of $\eta = 2.33\times 10^{12}$~cm$^2$~s$^{-1}$, which
corresponds to a Lundquist number, $S_L = v_a L/\eta$ $\approx 1.3 \times 10^4$ (where, $L$ is the characteristic length scale imposed by the current sheet, and $v_a$ is the Alfv{\'e}n speed at the base of the CS, both calculated at the initial time). The initial magnetic field configuration (Equations~\ref{eq:B_x}-\ref{eq:B_z}) is force-free, and the system is in initial hydrostatic equilibrium. This corresponds to a mechanical balance of the system at the initial state ($t=0$). To trigger the reconnection in the CS, we use a horizontal velocity perturbation ($v_p$) at multiple vertical locations which are concentrated in the CS region,

\begin{align}\label{eq:vp}
    v_p = 
    \begin{cases}
    -f \;v_a(y)\; 
\sin^2\left( 2 n\, \pi\, \frac{\displaystyle y}{\displaystyle L_y}\right)\; 
\frac{\displaystyle x}{\displaystyle \lambda}\,
\exp\left(-\frac{\displaystyle 1}{\displaystyle 20} 
\frac{\displaystyle x^2}{\displaystyle \lambda^2}\right); \text{for $y \geq 20$ Mm}\\
0; \text{for $y < 20$ Mm}
\end{cases}
\end{align}
Here, $L_y=40$~Mm is the vertical span of the CS and the Alfv{\'e}n speed is based on the initial density and magnetic field strength $\displaystyle{v_a(y) = \frac{B_0}{\sqrt{4\,\pi\,\rho(y,t=0)}}}$. We use  $n=3$ to pinch the CS through horizontal flows concentrated around three different heights ($y = 23.33,\; 30$, and $36.67$~Mm) and $f=0.05$, which gives a maximum $|v_p| \approx 37$~km~s$^{-1}$. This prescription of the velocity perturbation corresponds to a more intense pinching the higher along the $y$-axis. We use the  classical magnetic-field aligned heating rate associated with heat conduction as follows: $\nabla \cdot \left[\kappa_{||}\,{\bf b}\, ({\bf b} \cdot \nabla)\, T\right]$, with {\bf b} the unit vector along the magnetic field lines and ${\bf \kappa}_{\parallel} = 10^{-6}\, T^{5/2}$ erg cm$^{-1}$ s$^{-1}$ K$^{-1}$ is the Spitzer-type thermal conductivity. Thermal conduction is zero at the initial state of the system due to our isothermal atmosphere assumption, but it plays a role when the system deviates away from the isothermal condition as time evolves.

The MHD equations are solved numerically using a three-step, third-order
Runge-Kutta time integration method with the Van Leer flux limiter
\citep{1974:vanLeer} and Total Variation Diminishing Lax-Friedrichs (TVDLF)
flux scheme. We use the `splitting' option of the magnetic field variable ${\bf B}$ available in AMR-VAC, in which it is decomposed into a steady background term and a time-dependent deviation part \citep{2018ApJS..234...30X}. We use a symmetry condition (which copies the cell values of the physical boundaries to the ghost cells in a symmetric fashion) for $\rho$, $p$, $v_x$, $v_y$ and $v_z$ at the side boundaries. For the magnetic field components, we set $B_y = \pm B_0$ at the right and left boundaries respectively, and zero for the $B_x$ and $B_z$ components. At the top and bottom boundaries, $p$ and $\rho$ are set according to the hydrostatic assumption. All the velocity components at the top and bottom boundaries obey the symmetry condition, additionally by clipping the inflows to zero for the $v_y$ component at the top and bottom boundaries. $B_y$ and $B_z$ are symmetric (i.e., $\partial B_y/\partial y = 0$, $\partial B_z/\partial y = 0$) and $B_x=0$ at the top and bottom boundaries. Note that these boundary conditions maintain the solenoidality of the magnetic field at all boundaries. We also use the divergence of $\bf B$ cleaning with parabolic diffusion method in the whole domain \citep{Keppens:2003, keppens2023}. 

\begin{figure}
    \centering
    \includegraphics[width=0.48\columnwidth]{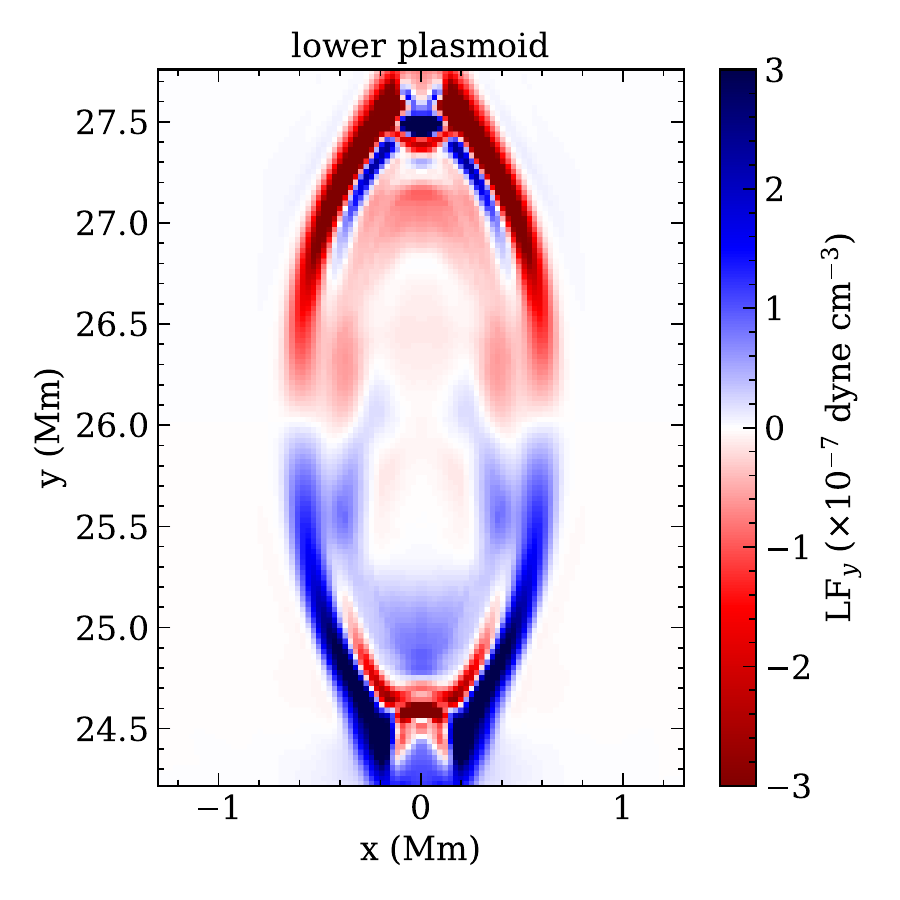}
    \includegraphics[width=0.48\columnwidth]{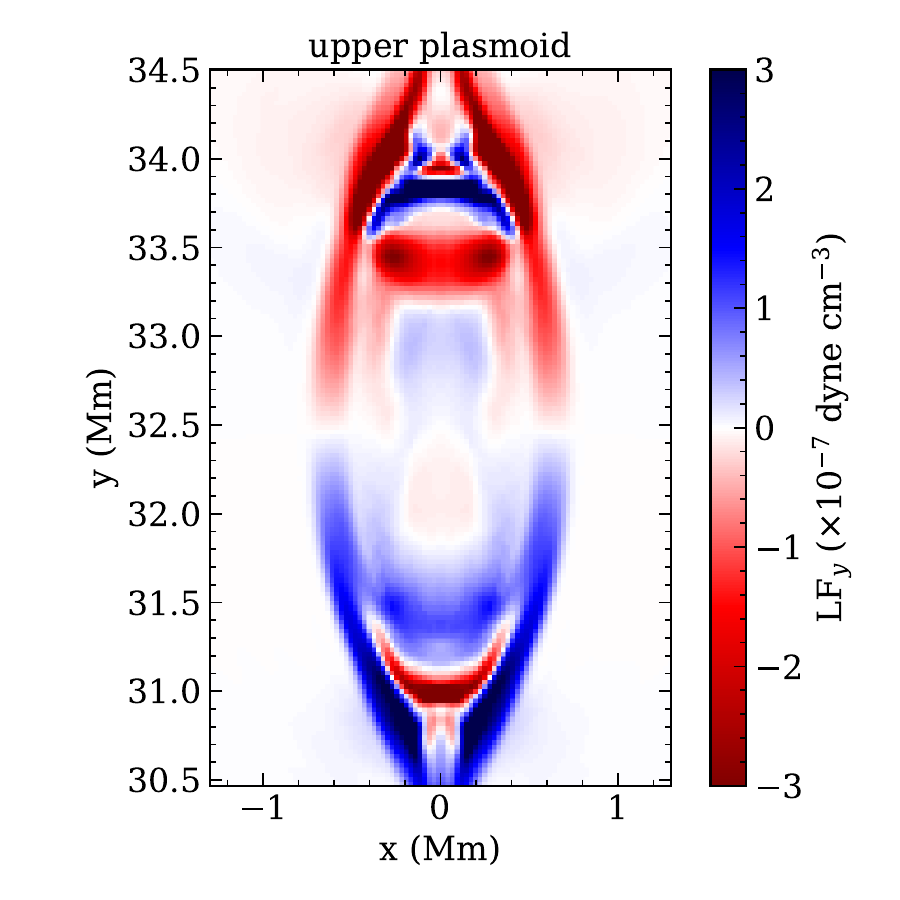}
    \caption{Spatial distribution of the vertical component of Lorentz force covering the lower (left panel) and the upper (right panel) plasmoids at $t=81.58$ s.}
    \label{fig:LFy-plasmoids}
\end{figure}

\begin{figure*}
    \centering
    \includegraphics[width=0.33\textwidth]{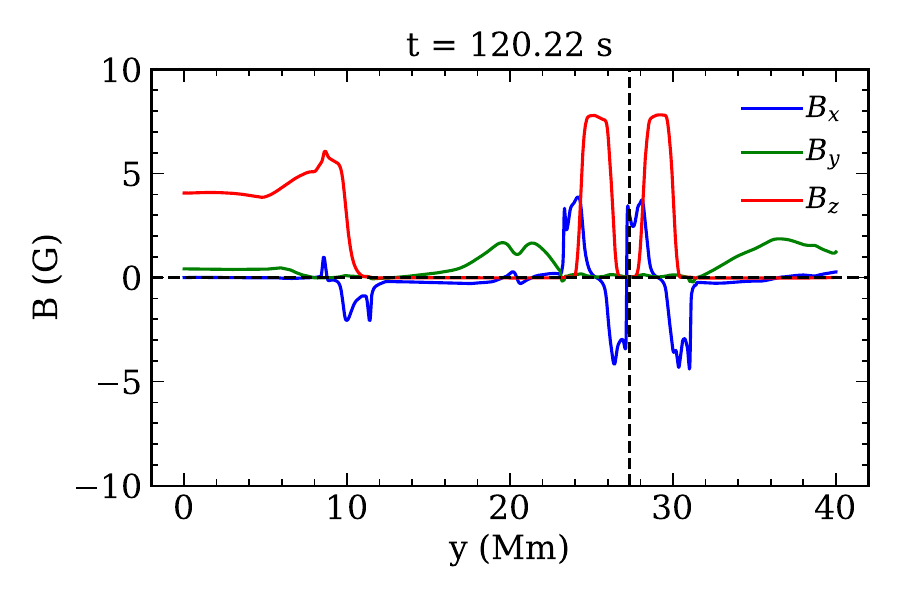}
    \includegraphics[width=0.33\textwidth]{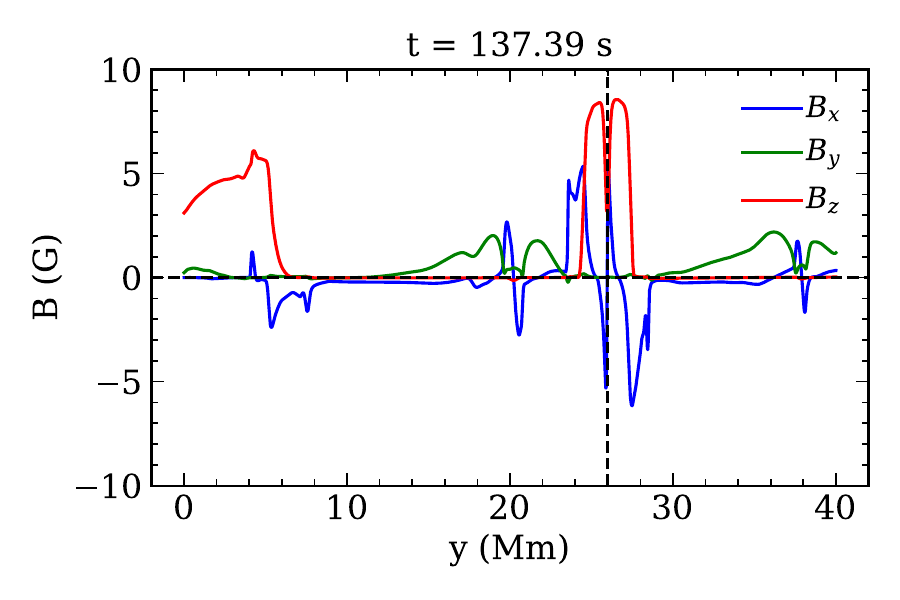}
    \includegraphics[width=0.33\textwidth]{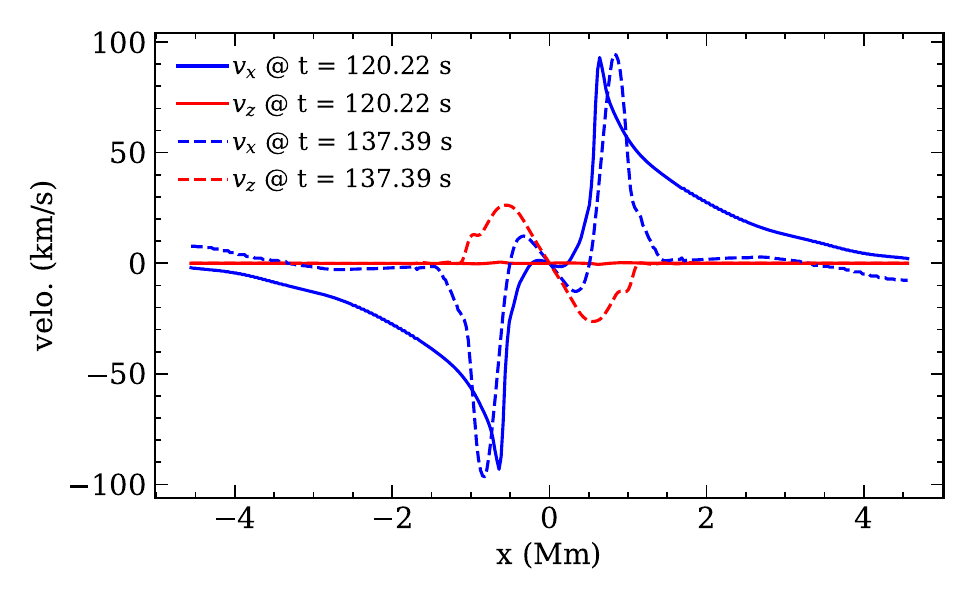}
    \caption{The left and middle panels show the variation of the magnetic field components at $x=0$ along the $y$ direction at $t=120.22$ and $137.39$ s respectively. The black horizontal dashed lines mark the zero field components. The vertical black dashed lines are located at $y=27.34$ and $26$ Mm (at $x=0$) respectively for the left and middle panel figures. The right panel shows the distribution of the  $v_x$ and $v_z$ components along a horizontal axis situated at those values of $y$ at $t=120.22$ and $137.39$ s respectively.}
    \label{fig:2d-3d-reconnection-transition}
\end{figure*}

\begin{figure}
\centerline{\hfill \includegraphics[width=0.5\textwidth]{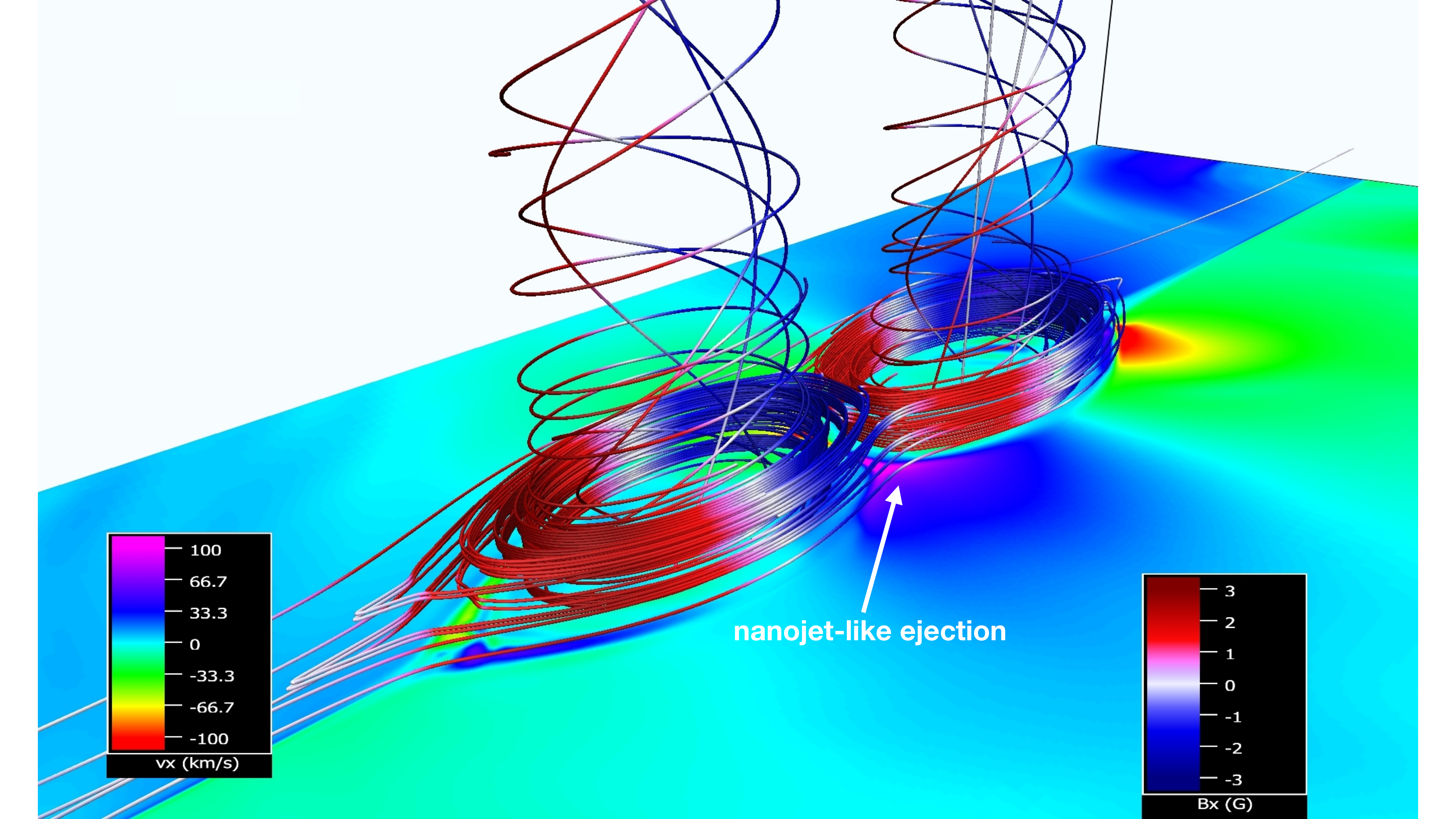}}
\hfill
\caption{The magnetic field configuration during the initial, 2D-like, phase of the reconnection associated with the plasmoid merger ($t=120.22$~s). 
A colormap is drawn for $v_x$; the field lines are colored according to the value of $B_x$. The appearance of a small nanojet-like feature is marked by a white arrow.}
\label{fig:reconnection_2D}
\end{figure}

\begin{figure*}[hbt!]
\centering
\hbox to \hsize{\hskip -0mm
\includegraphics[width=0.57\textwidth]{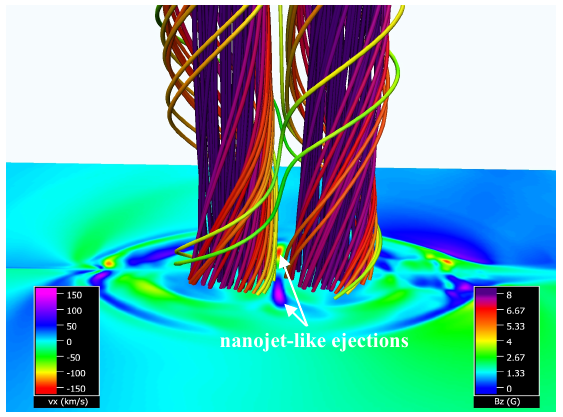}\hskip 2mm
\vbox{\hsize 0.38\textwidth
\centerline{\ \hskip 3mm\includegraphics[width=0.38\textwidth]{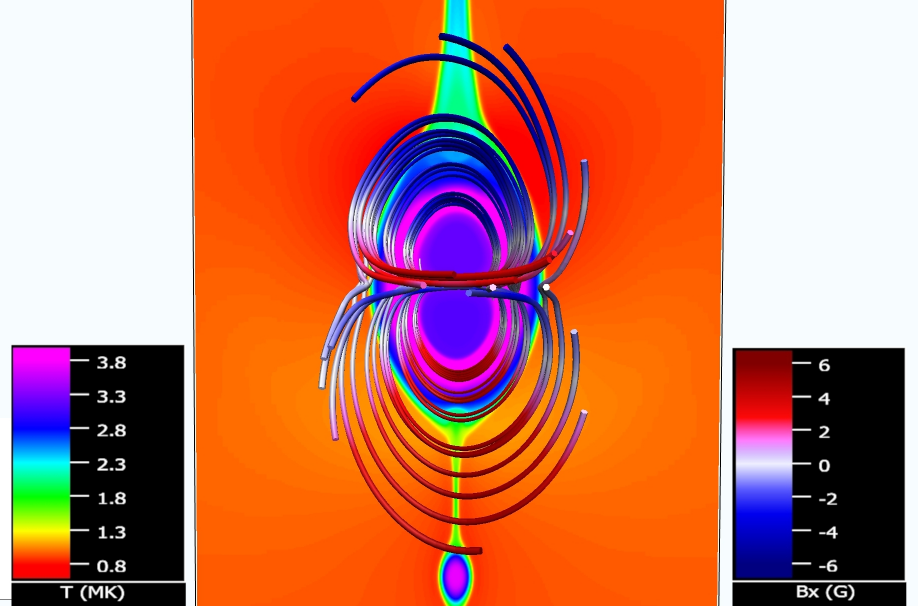}}\vskip 3pt
\includegraphics[width=0.40\textwidth]{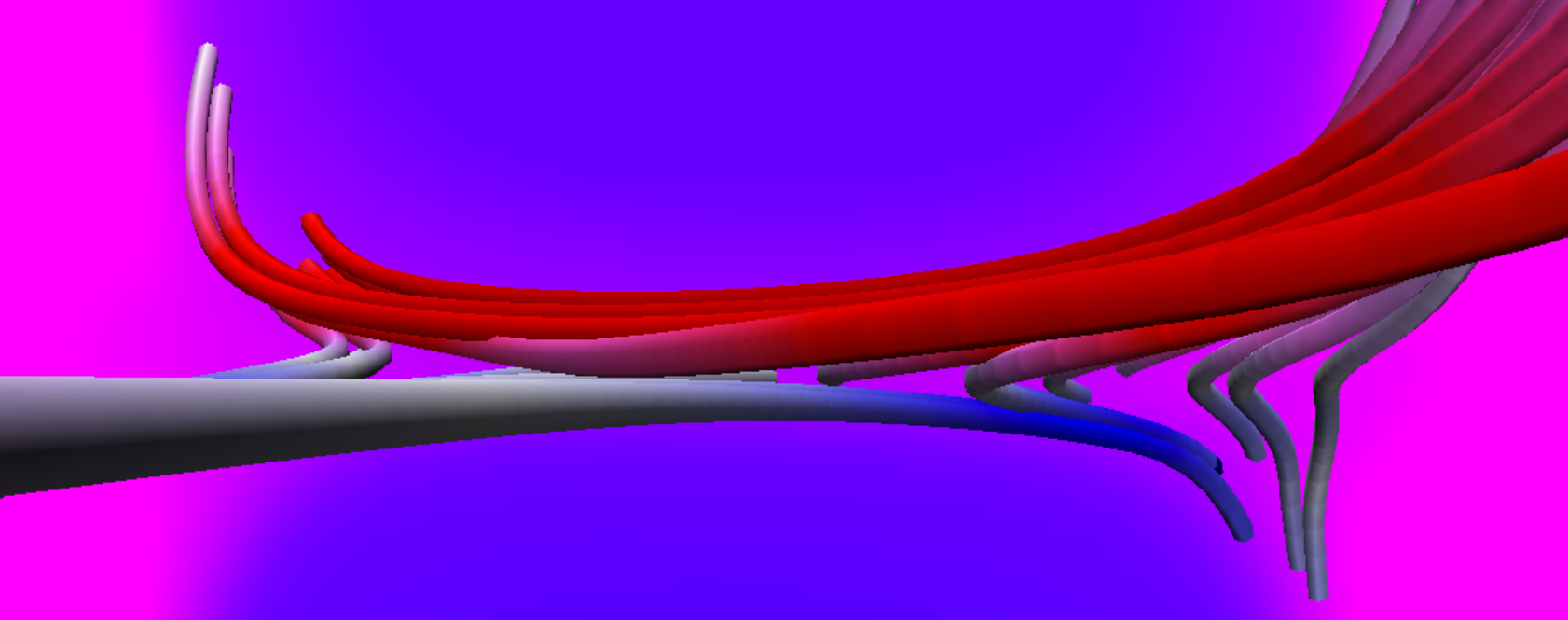}
}}
\caption{Selected field lines at $t=139.53$ s during the process of merging of the two plasmoids, showing details of the 2D reconnection with a guide field process between them. Left panel: the background map is for $v_x$ (in km s$^{-1}$); the color along the field lines corresponds to the $B_z$ field component. Top right panel: top view of the same configuration but with field lines now traced from the transverse current sheet where the merger is taking place. The background map corresponds to $\log T$. Bottom right panel: Same as the top right panel, but for a selected region highlighting the interface between the upper and lower plasmoids.
}
\label{fig:3dFL}
\end{figure*}

\begin{figure}[hbt!]
    \centering
    \includegraphics[width=0.5\textwidth]{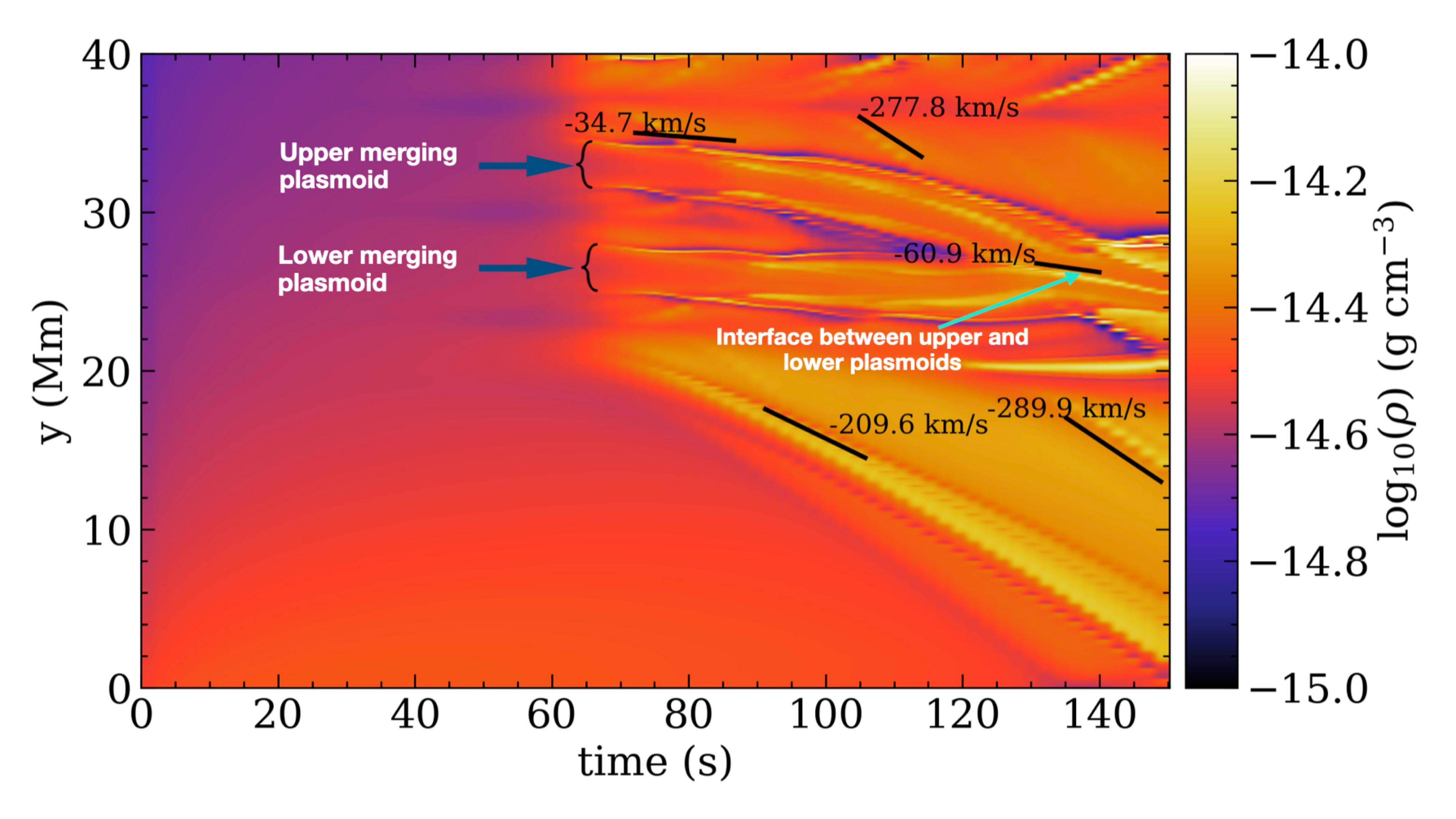}
    \caption{Time-distance map of density at vertical (along $y$) cut at $x=0$. The solid lines for estimating the slopes at the selected locations of the time-distance map are marked in black. The vertical extent of the upper and lower plasmoids is marked with braces and pointed by the blue arrows, the interface between the plasmoids is marked with a cyan arrow.
        \label{fig:rho-TD}}
\end{figure}

\begin{figure*}
    \centering
    \includegraphics[width=0.3\textwidth]{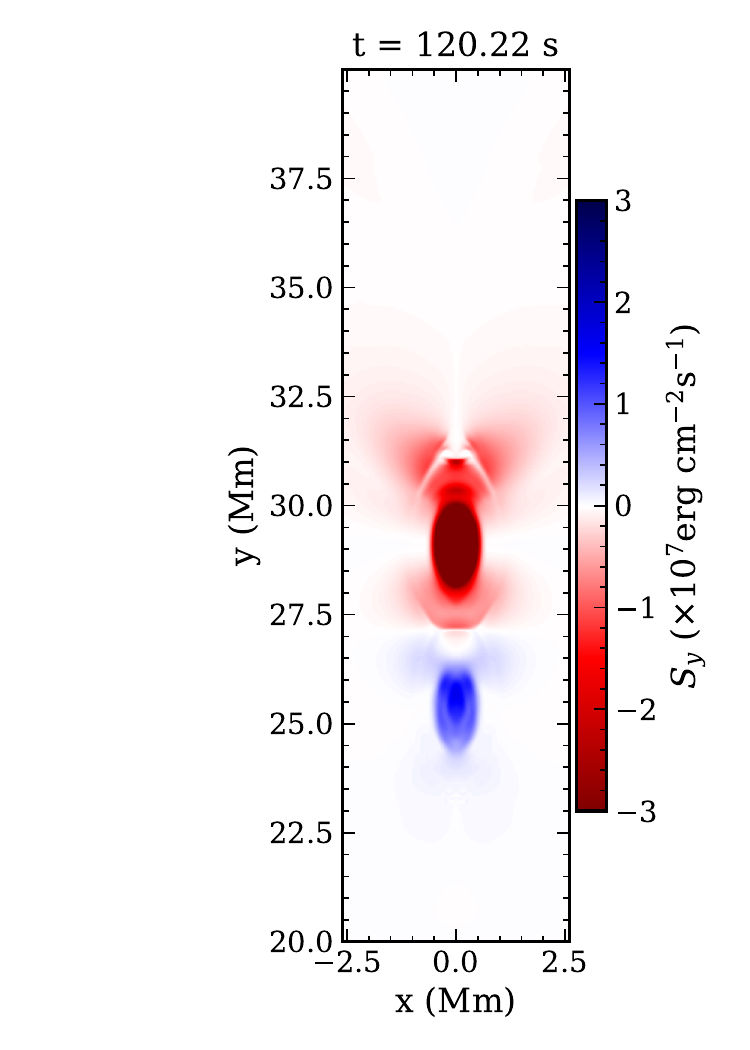}
    \includegraphics[width=0.3\textwidth]{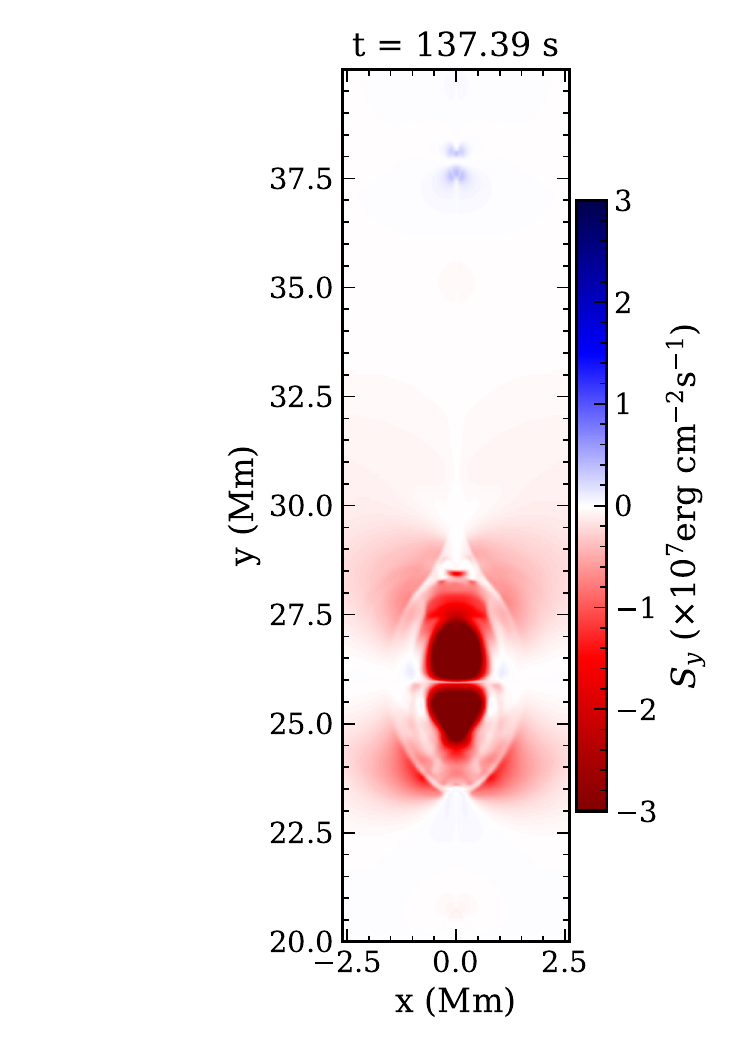}
    \includegraphics[width=0.33\textwidth]{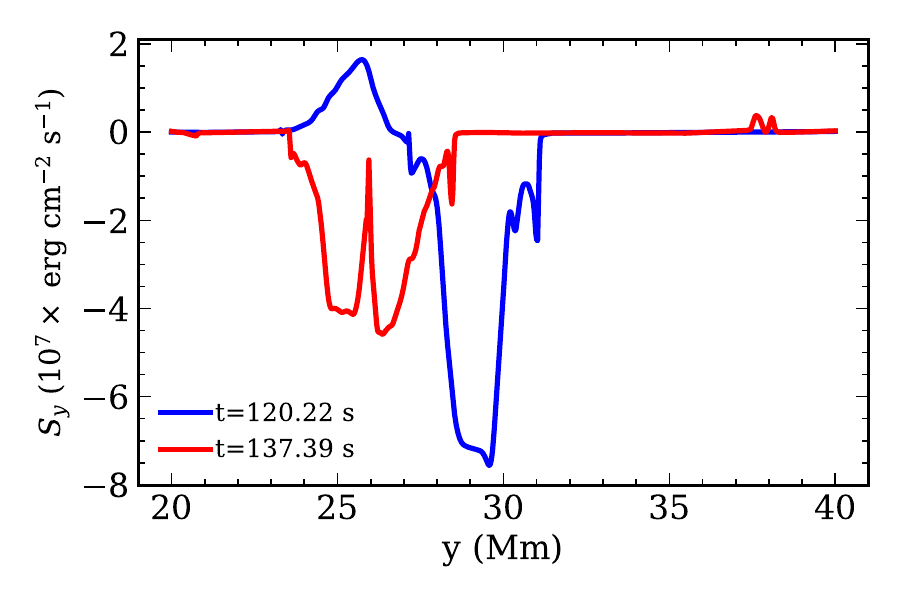}
    \caption{The left and middle panels show the spatial distribution of the vertical Poynting flux vector, $S_y$, in the early and advanced merging stages of the plasmoids, at $t=120.22$ and $t=137.39$ s, respectively. The right panel shows the vertical distribution of $S_y$ at $x=0$ at $t=120.22$ and $t=137.39$ s.}
    \label{fig:Sy}
\end{figure*}

\begin{figure*}[hbt!]
    \centering
    \includegraphics[width=1\textwidth]{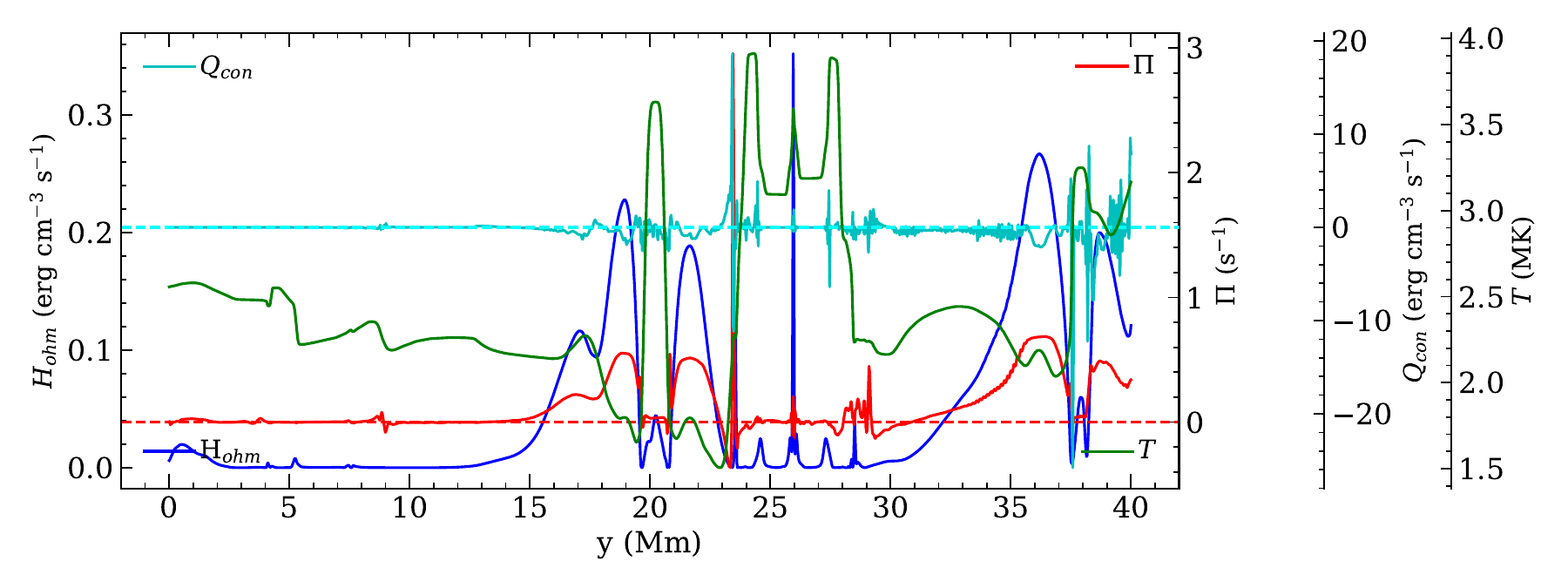}
    \caption{Vertical distributions of Ohmic heating rate ($H_{ohm}$), expansion/contraction rate ($\Pi$), heating rate by thermal conduction ($Q_{con}$), and temperature ($T$) at $x=0$ for time $t=137.39$ s. The red and cyan horizontal dashed lines represent the zero markers for $\Pi$ and $Q_{con}$ respectively.} 
    \label{fig:Temp-cause}
\end{figure*}

\begin{figure*}
    \centering
    \includegraphics[width=0.3\textwidth]{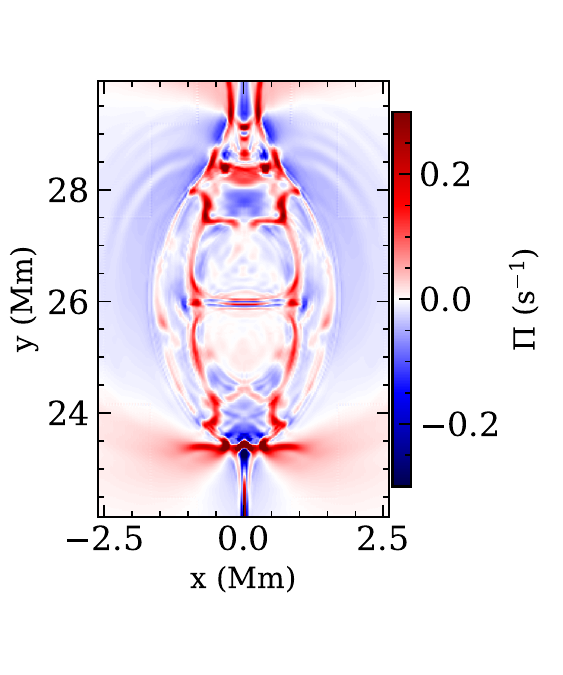}
    \includegraphics[width=0.29\textwidth]{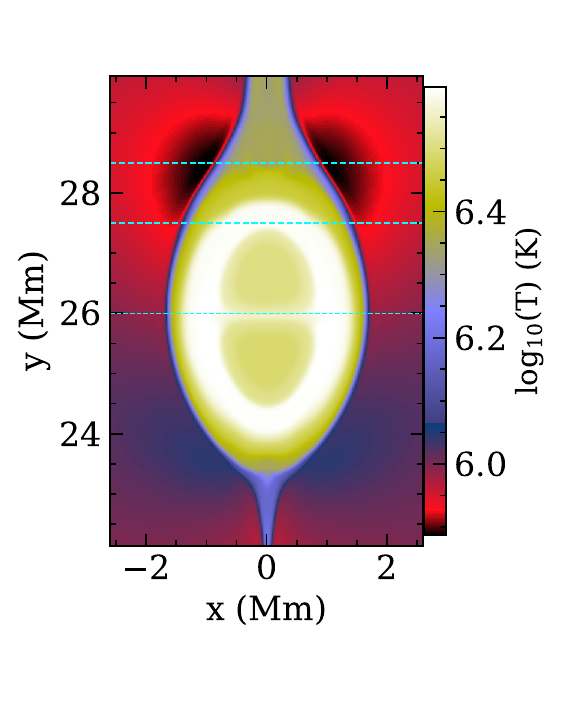}
    \includegraphics[width=0.3\textwidth]{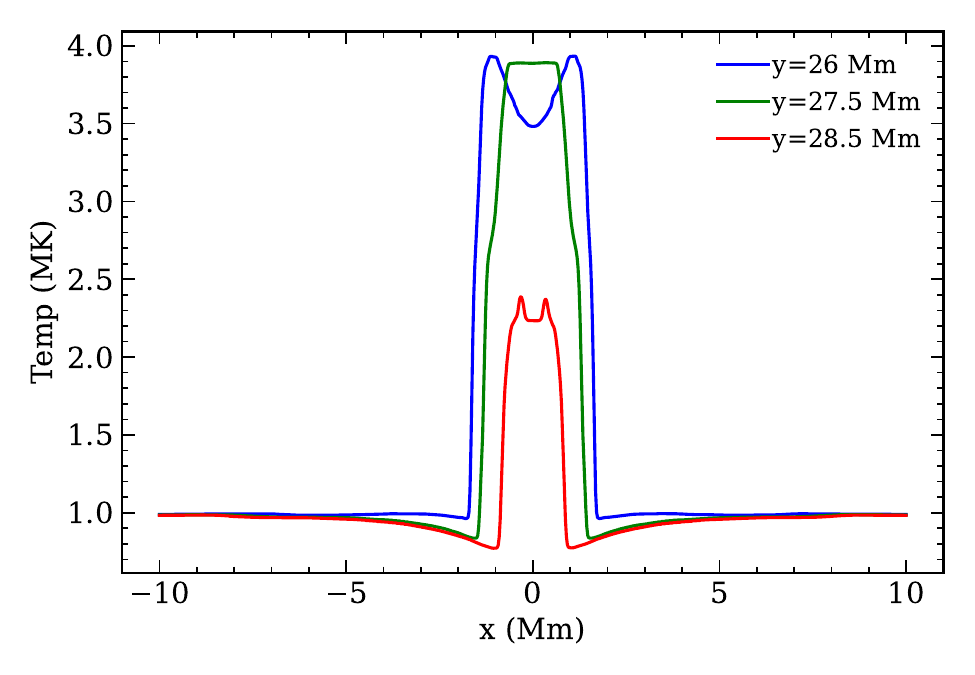}
    \caption{The left and middle panels show the expansion/contraction rate and the temperature distribution, respectively, in the region of the merging plasmoids at $t=137.39$ s. The dashed horizontal (cyan) lines in the middle panel mark the levels $y=26$, $27.5$, and $28.5$ Mm. Right panel: Temperature distribution along the horizontal direction for the three $y$-levels marked by the dashed lines in the middle panel. The minimum temperature along the $y=28.5$~Mm line (red curve in the right panel) is $0.8$~MK.}
    \label{fig:pi_Te_xy}
\end{figure*}

\section{Results and analysis}\label{sec:results}
\subsection{Formation of plasmoids and their dynamics} \label{subsec:plasmoid}

The velocity perturbation in the CS squeezes the opposite-polarity magnetic field lines, which leads to the initiation of magnetic reconnection at the central vertical axis ($x=0$), and makes the CS unstable. In fact, the initial pulses in $v_x$ coming from either side of that axis collide against each other soon after $t=0$. They then bounce back and move away from the y-axis as weak magnetosonic shocks, leaving the domain through the side boundaries at $t\approx 40$~s. The initial compression of the CS is enough to start the tearing mode instability in it, which leads to the gradual development of plasmoids, as shown in the upper panel of Figure \ref{fig:combined} and the associated animation, available online. We notice the development of two primary plasmoids at $y \approx 33$ and $27$~Mm at $t\approx 40$~s; we shall call them the upper and lower plasmoid, respectively, and the discussion in the rest of the paper will be focused onto them. The growth of magnetic islands formed at different heights is not identical: pinching due to the velocity pulse, $v_p(y)$ (see equation \ref{eq:vp}) has velocities of $36.8$~km s$^{-1}$, $35.7$~km s$^{-1}$, and $34.5$~km s$^{-1}$ at $y=36.67$, $30$, and $23.33$~Mm, respectively; this causes more intense reconnection the higher along the CS. On the other hand, the reconnection taking place at those three levels sweeps plasma with purely planar field (i.e., with zero $B_z$ component) from the sides of the simulation into the central axis and toward the plasmoids. At the same time, the $B_z$ field contained in the CS is shuffled toward the plasmoid interior: the plasmoids actually become solenoids with a guide field pointing along the symmetry axis $z$, surrounded by plasma with purely planar field. The energy released via Ohmic heating at the CS, specially at the reconnection sites, leads to a temperature increase, which is the reason for the hot plasma [$\log T(K) \sim 6.5$] apparent in the fourth column of Figure~\ref{fig:combined}. Additionally to the two major plasmoids, there is a plasmoid-like structure starting at $y\approx 20$~Mm that travels quickly toward the lower boundary. It is generated right below the lowest pulse at the initial time, located at $y=23.33$~Mm. Because of the initial pulse, the reconnection rate above that plasmoid is larger than below it, leading to a marked imbalance of the vertical component of the Lorentz force and fast downward motion. A similar feature was obtained in the experiment of \citet{Popescu_Keppens_2023}, who had a single velocity pulse in the whole domain.  

Figure~\ref{fig:forces} illustrates the distribution of the vertical component of the Lorentz force (LF), the gas pressure gradient force ($-\frac{dp}{dy}$), and the gravity force ($\rho g$),  along the vertical line at $x=0$ for $t=81.58$~s,
which is the time shown in the top row in Figure~\ref{fig:combined}. It is clear that the vertical component of the LF dominates over the other two forces. In Figure~\ref{fig:LFy-plasmoids}, we present the spatial distribution of the vertical component of the LF in  the regions occupied by the lower and upper plasmoid at $t=81.58$ s; the location of those plasmoids in the $y$-direction is also highlighted with gray shading in Figure \ref{fig:forces}. The red versus blue area distribution in either plasmoid appears to have a rough top-bottom symmetry. To determine the sign of the vertical acceleration being imparted to the plasmoid by the LF, we employ binary masking to identify pixels with unsigned vertical component of LF above a threshold value (namely, $1 \times 10^{-8}$ dyne cm$^{-3}$), which, according to Figure~\ref{fig:LFy-plasmoids}, can serve as a proxy for the location of the plasmoid. The integration of the y-component of the LF over the plasmoid region yields $-7.3\times 10^7$~dyne~cm$^{-1}$ (upper plasmoid) and $-1.3\times 10^7$~dyne~cm$^{-1}$ (lower plasmoid). This corresponds to a fraction $2.3$\% and $0.49$\%, respectively, of the integrated unsigned vertical component of the LF. Therefore, in spite of the apparent symmetry, the net LF points downward in both plasmoids. Dividing the integrated (signed) LF by the integrated density of the plasmoid, we obtain an estimate for the center-of-mass acceleration due to the LF, namely, $-6.9$~km~s$^{-2}$ (upper plasmoid) and $-1.3$~km~s$^{-2}$ (lower plasmoid). This difference in the acceleration values (factor $5.3$) is the reason why the upper plasmoid approaches the lower one when descending, a process that finally leads to their coalescence, as shown in the middle row of Figure \ref{fig:combined} at around $t=120$~s and also in the animation accompanying that figure. Concerning the reason for the sign of the net vertical Lorentz force, this is most probably a result of a {\it melon-seed} expulsion process \citep{Schluter_1957}: as already said, the converging flows that lead to the creation of the plasmoids are stronger the higher along $y$. This suggests that the reconnection at the top of either plasmoid is more intense than just underneath them, and we expect the vertical component of the LF to be stronger at the upper end of the plasmoid than in the lower one, as can also be seen in Figure~\ref{fig:LFy-plasmoids}. 

The two plasmoids gradually merge into a single, larger one that contains two baby solenoids within it (see the left panel of the bottom row of Figure \ref{fig:combined}). The merger is achieved through field line reconnection across the interface between the two plasmoids; in fact, this reconnection process can be seen to evolve in time from basically purely 2D reconnection to 2.5D reconnection, i.e., 2D reconnection with a guide field along the ignorable coordinate. As seen above, the initial configuration of the plasmoid is that of a solenoid-like twisted flux rope with the guide field pointing in the invariant $z$-direction and surrounded by field contained in the $xy$ plane. When first meeting each other at, say, $t=120.22$~s, the magnetic field lines that get into contact point essentially along the $x$-direction but in opposite senses. This can be seen in Figure~\ref{fig:2d-3d-reconnection-transition}, left panel, which shows the three components of the magnetic field along the $y$-axis at $x=0$. The result is a process of standard 2D reconnection. At the time of the central panel ($t=137.39$ s), instead, the cores of the solenoids have already been brought into contact: one sees that now 2.5D reconnection, whereby the reconnecting component is still $B_x$, but now there is a dominant guide field along the $z$-direction so that the mutual angle between the reconnecting elements is far from $180^\circ$. Concerning the velocities, additional information of interest can be gained through the corresponding profiles of $v_x$ and $v_z$ along the horizontal axis that cuts across the reconnection site (Figure~\ref{fig:2d-3d-reconnection-transition}, right panel). During the initial phase ($t=120.22$~s), there is a marked outflow pattern in the $x$-direction, clearly corresponding to the reconnection outflows of the initial, 2D-like configuration; these outflows can be seen to extend for some $4$-Mm on either side of the reconnection point; by $t=137.4$~s, we see that the reconnection outflows still have velocities of order $100$~km~s$^{-1}$ for $v_x$ but, now, incipient outflows in the $z$-direction are also apparent. Additionally, the spatial extent along $y$ has become much narrower: this can be seen in the accompanying animation (entitled 'v$_x$.mp4'), which is available online. 

To appreciate the magnetic field line configuration at the reconnection site, we provide in Figures \ref{fig:reconnection_2D} and \ref{fig:3dFL} a 3D view of the field line exchange taking place at the interface. In Figure~\ref{fig:reconnection_2D} we show the configuration during the initial, 2D-like phase of the reconnection, at $t=120.22$~s. We can confirm that the field lines in the periphery of the plasmoids basically lie on the $xy$-plane and, also, that, at the contact point, field lines subtending a mutual angle close to $180$~deg (see the colors along the field lines, which correspond to $B_x$) are being reconnected and hurled sideways in the $x$-direction: the colormap on the $xy$ plane for $v_x$ shows two concentrated outflows with high speeds of approx ~$\pm 100$~km~s$^{-1}$ originating in the reconnection site. As time advances, the central regions of the two plasmoids get increasingly close to each other; their field lines have a predominant $z$-component, with $B_z >0 $; their $B_x$ component, instead, points in opposite directions and this maintains the transverse current sheet apparent at $t=120.2$~s and $t=137.4$~s in Figure~\ref{fig:combined}. The hybrid field lines shown in Figure~\ref{fig:3dFL} (i.e., those which have just reconnected and thus link the two plasmoids), drawn for $t=139.5$~s, while still having $B_z > 0$, are hurled away from the reconnection site with marked curvature, and hence, large Lorentz force and speeds, as confirmed through the background color map for $v_x$. This is apparent in the left panel; here, the field lines were all traced from points in a box that includes the central part of the merging plasmoids and the transverse CS; the color along the field lines corresponds to the value of $B_z$. We see how the field lines from either plasmoid that pass near the CS are mutually inclined; one can also see a few just reconnected field lines (like, e.g., the green-yellow one) which now link the two merging plasmoids. In the right panel, a top view is shown, now on the background of a color map for $\log T$ and with field lines traced from the neighborhood of the transverse CS. By magnifying the central part of the figure (see the subpanel on the bottom right), one can see the curvature of the just reconnected field lines, which are being hurled in the $\pm x$-direction. The plasma gets strongly heated in the reconnection process; the background map clearly shows how this plasma moves along the resulting field lines, so that the periphery of the merged plasmoid has temperatures of several MK.

Space-time information on the formation of the various plasmoids and their associated dynamics is shown in Figure \ref{fig:rho-TD}. Here we compute a time-distance (TD) map of the plasma density along the vertical line at $x=0$. We estimate the speed of the plasmoids by tracking the enhanced density region in the TD map corresponding to the plasmoid motions, and calculating their respective slopes. Most of the identified plasmoids have speeds within the range $\approx 35-290$~km~s$^{-1}$. The primary upper and lower plasmoids that appear at around $t=80$~s are approximately $3$~Mm in size along the vertical direction (marked by the black braces); the downward speed of the upper plasmoid is around $35$~km~s$^{-1}$, whereas the lower plasmoid moves much more slowly and its global velocity cannot be determined with the same level of accuracy as for the others. In fact, when the upper plasmoid starts to merge with the lower one at around $t=120$~s, the interface between the upper and lower plasmoids is pushed downward at a speed of around $60$ km s$^{-1}$. Around $t=137$ s, when the merging is complete, the size of the merged plasmoid starts to grow vertically and ends up reaching a vertical size of $\approx 8$ Mm, and interacting with another (nearly stationary) secondary plasmoid (forms at $y\approx 20$ Mm) at the end of the simulation. The downward moving plasmoid that appears at $y\approx 20$ Mm at around $t=64$ s gains a high speed of around $210$ km s$^{-1}$ as marked by the white arrow in Figure~\ref{fig:rho-TD}, due to the imbalance in the vertical component of the Lorentz force, as explained above.

\begin{figure}[hbt!]
    \centering
    \includegraphics[width=0.489\textwidth]{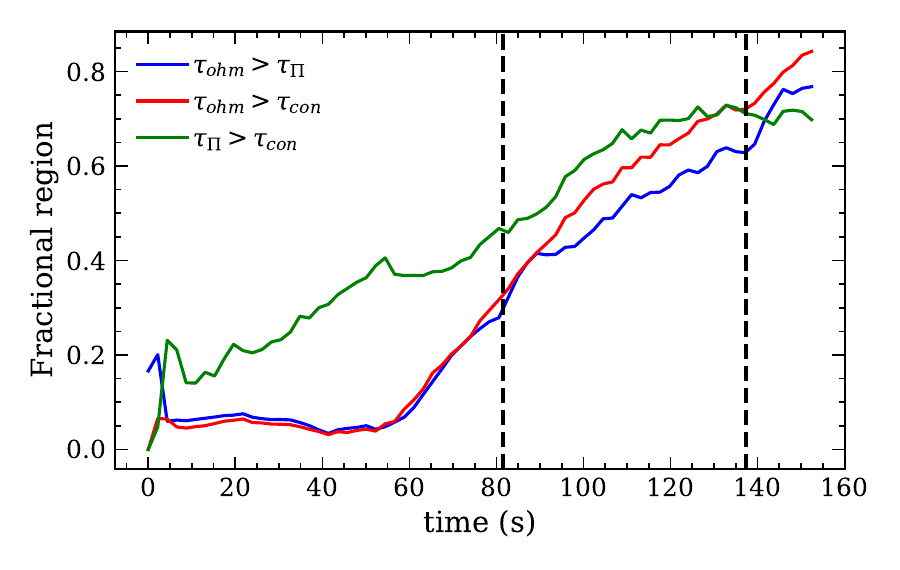}
 \caption{Temporal variation of the fractional counts of pixels that satisfy the conditions indicated in the legend. The vertical dashed lines are marked at $t=81.58$ and 137.39 s.} 
    \label{fig:timescales-evolution}
\end{figure}

\begin{figure*}[hbt!]
    \centering
    \begin{subfigure}[b]{1\textwidth}
        \centering
        \includegraphics[width=1\textwidth]{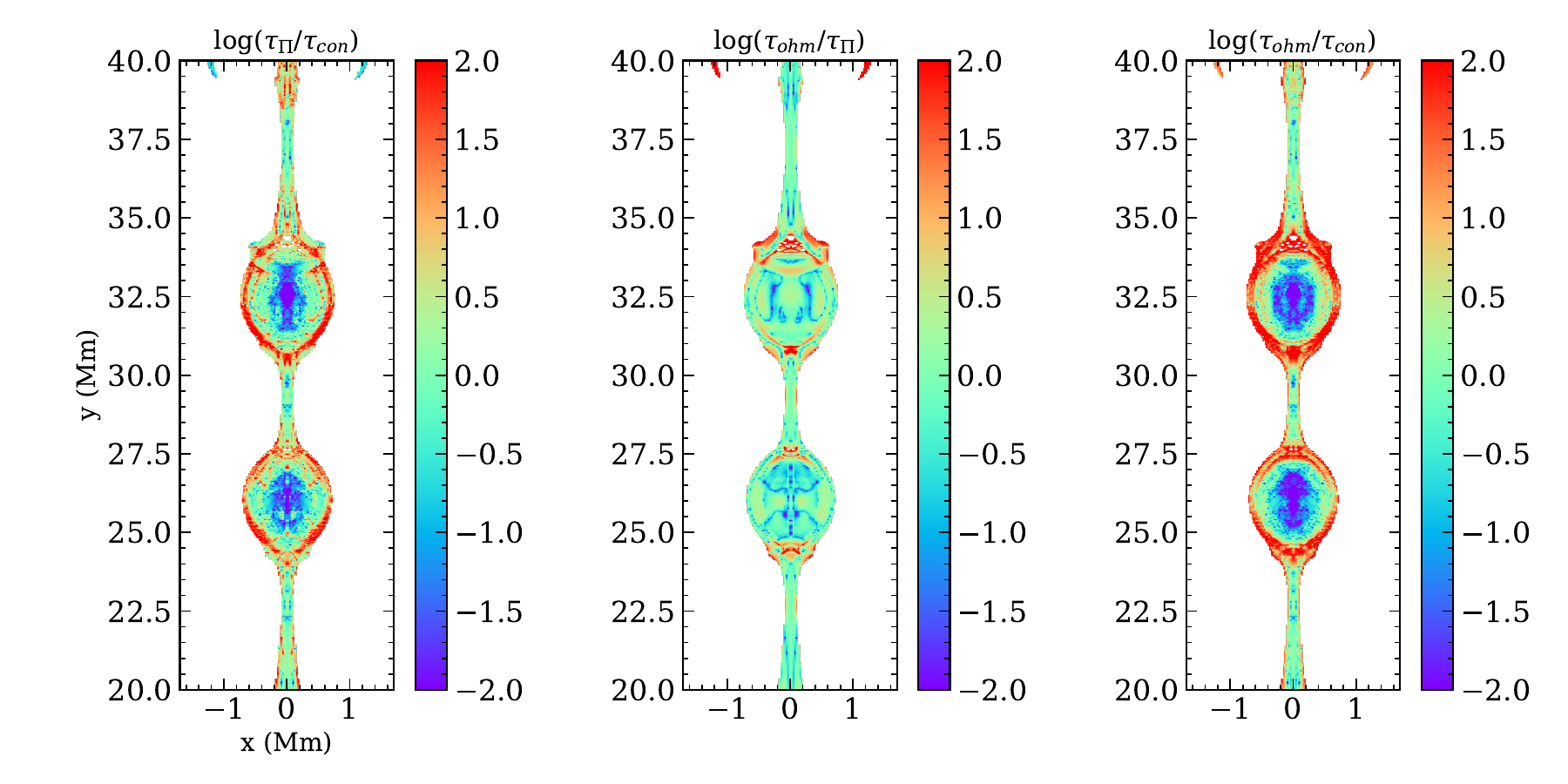}
        \caption{$t=81.58$ s}
        \label{fig:log_tau_sub1}
    \end{subfigure}
    \hfill
    \begin{subfigure}[b]{1\textwidth}
        \centering
        \includegraphics[width=1\textwidth]{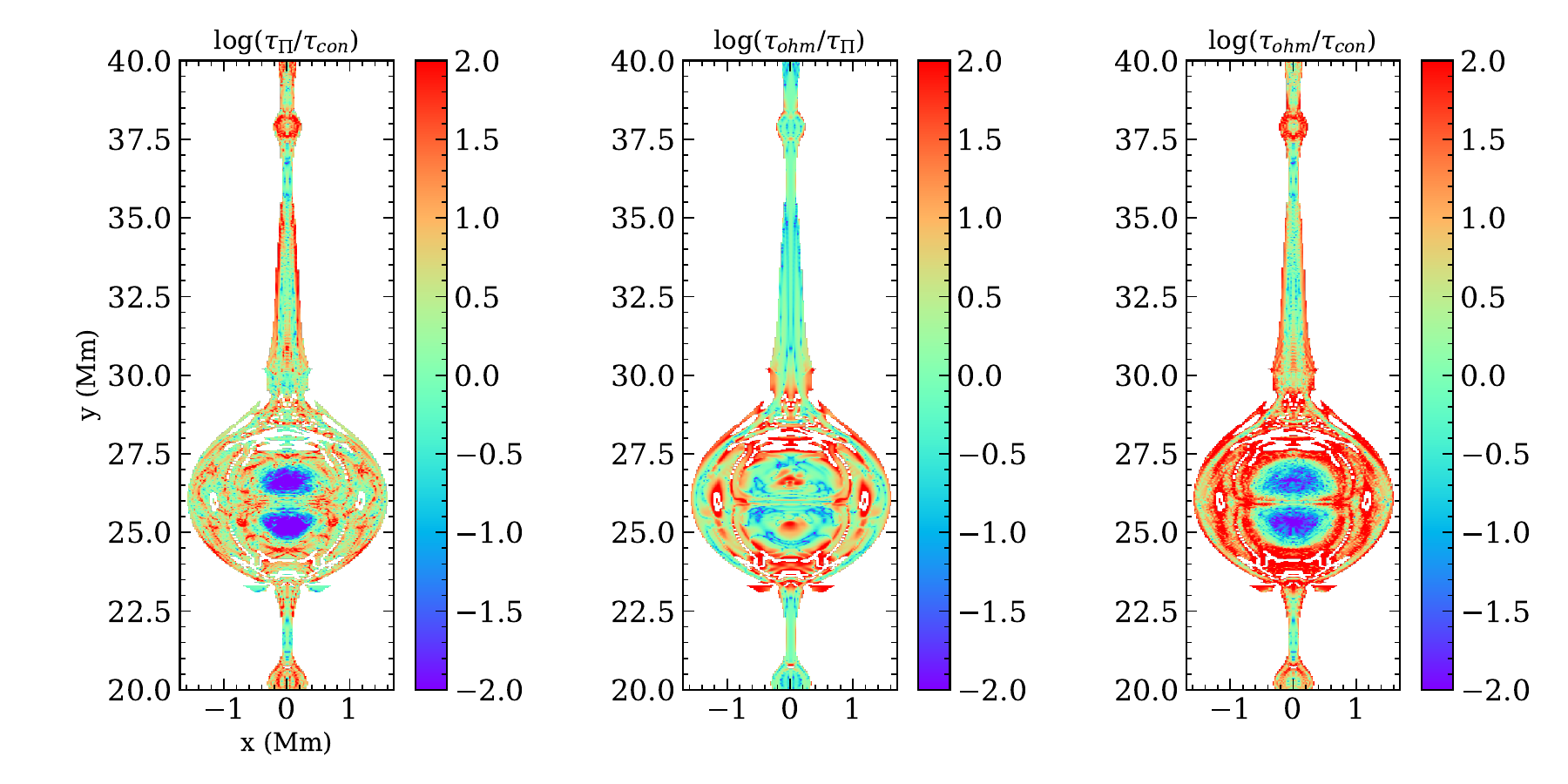}
        \caption{$t=137.39$ s}
        \label{fig:log_tau_sub2}
    \end{subfigure}    
       \caption{The top and bottom rows represent the \textit{``TECZ"} regions (that satisfy the criterion $|J|>J_{th}$) for the pre-merger and advanced merging states at $t=81.58$ (top row) and 137.39 s (bottom row) respectively. The timescale ratios $\tau_\Pi/\tau_{con}$, $\tau_{ohm}/\tau_\Pi$, and $\tau_{ohm}/\tau_{con}$ are shown in the left, middle, and right columns, respectively, using logarithmic scales.} \label{fig:timescales-compare}
\end{figure*}

\begin{figure*}[hbt!]
    \centering
    \begin{subfigure}[b]{1\textwidth}
        \centering
        \includegraphics[width=1\textwidth]{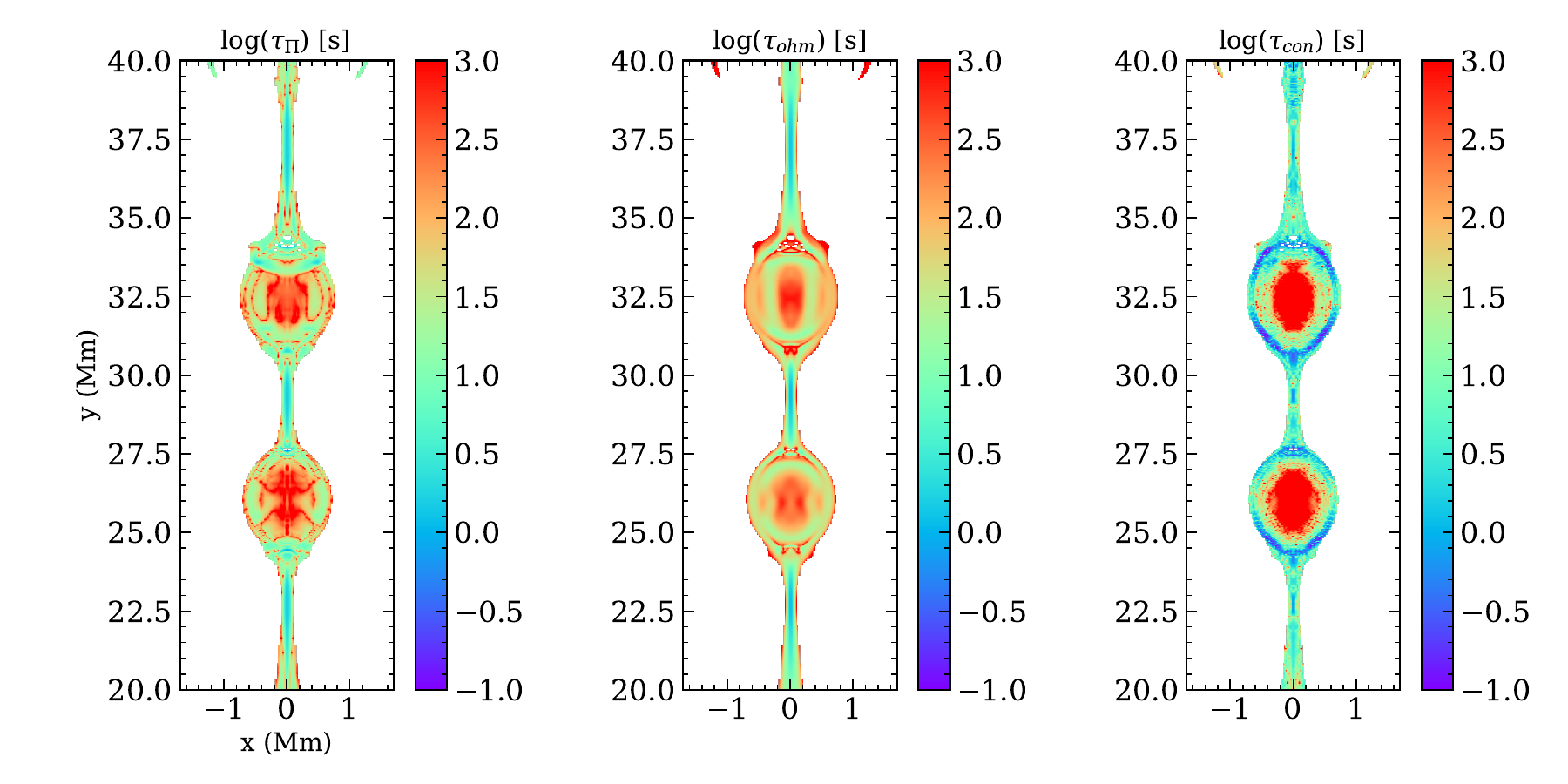}
        \caption{$t=81.58$ s}
        \label{fig:sub1}
    \end{subfigure}
    \hfill
    \begin{subfigure}[b]{1\textwidth}
        \centering
        \includegraphics[width=1\textwidth]{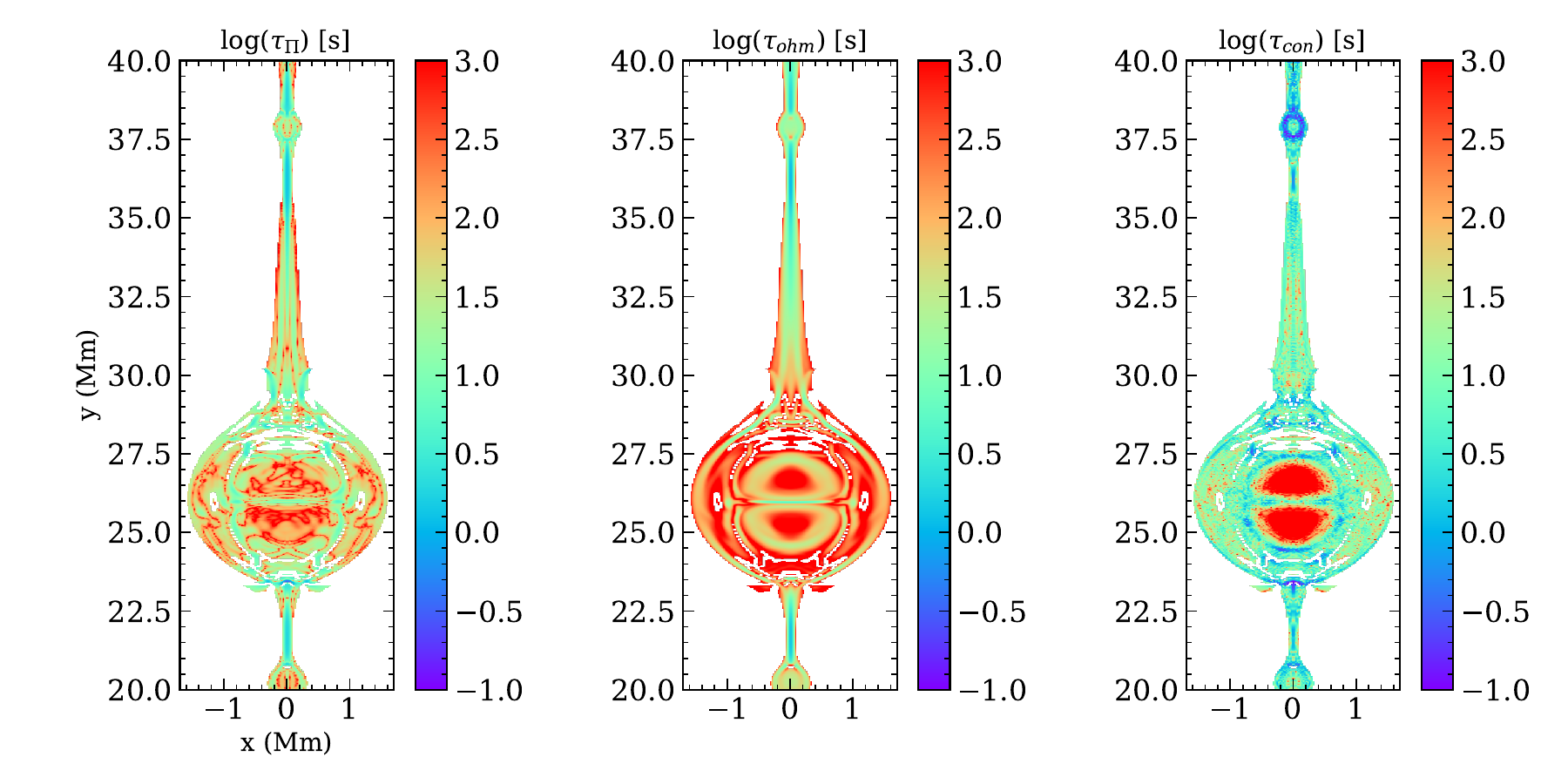}
        \caption{$t=137.39$ s}
        \label{fig:sub2}
    \end{subfigure}    
    \caption{Same as Figure \ref{fig:timescales-compare}, but showing the different characteristic time scales: $\tau_{\Pi}$ (left column), $\tau_{ohm}$ (middle column), and $\tau_{con}$ (right column).}
    \label{fig:timescales}
\end{figure*}

\subsection{Energetics and thermodynamic evolution}\label{subsec:thermodynamics}
To quantify the magnetic energy flux, we calculate the vertical component of the Poynting flux vector, 
\begin{align}\label{eq:Sy}
    S_y = \frac{1}{4\pi} [v_y\,(B_x^2+B_z^2)-B_y\,(v_x\,B_x+v_z\,B_z)],
\end{align} 
which is likely to be the main component of the energy transport in a low plasma-$\beta$ medium like the solar corona. We estimate $S_y$ before and after the plasmoids are fully merged as shown in Figure \ref{fig:Sy}. At the earlier time (left panel and blue curve in the right panel), we notice that the energy transport by the upper plasmoid, which is moving downward, is more intense than that corresponding to the lower plasmoid; in the more advanced time, after merging (middle panel and red curve in the right panel), the energy transport is dominated by the resulting downward-moving plasmoid. To further analyze the energy balance, we consider the equation for the internal energy. Since the calculation in this paper is for the simplest ideal gas with no ionization/recombination processes, we can express that equation in terms of the logarithmic change of the temperature in time, as follows:
\begin{align}\label{eq:energy_temperature}
    \frac{D\, \log T}{Dt} = \frac{H_{ohm}}{\epsilon} \;+\; \frac{Q_{con}}{\epsilon}\; - (\gamma-1)\, \nabla \cdot{\bf v} ,
\end{align}
where $D/Dt$ is the Lagrange (or material) derivative, $\epsilon$ the internal energy of the gas per unit volume, and $\gamma=5/3$ is the customary ratio of specific heats. In Eq.~\ref{eq:energy_temperature} we recognize the two possible entropy sources in the system, namely Ohmic heating ($H_{ohm}$) and heat deposition by heat conduction ($Q_{con}$) given by

\begin{align}
   & H_{ohm} = \eta J^2,\\
\noalign{\vskip 1mm}
   & Q_{con} = \nabla \cdot \left[\kappa_{||}\,{\bf b}\, ({\bf b} \cdot \nabla)\, T\right];\\ 
\end{align}
additionally, we shall call $\Pi$ the remaining term:
\begin{equation}\label{eq:Pi}
\Pi = -(\gamma-1) \,\nabla \cdot {\bf v} \,,
\end{equation}
which corresponds to the change in temperature because of expansion/contraction even in the absence of entropy sources. Equation~\ref{eq:energy_temperature} expresses the inverse of the timescale for the change of temperature as a combination of the inverse values of the timescales associated with Ohmic heating ($\tau_{ohm}$), heat conduction ($\tau_{con}$) and expansion/contraction ($\tau_\Pi$) each with its sign, with the timescales defined as:
\begin{align} \label{eq:t_ohm}
    & \tau_{ohm} = \epsilon/H_{ohm},\\
\noalign{\vskip 2mm}
 \label{eq:t_con}
    & \tau_{con} = \epsilon/|Q_{con}|,\\ 
\noalign{\vskip 2mm}\label{eq:t_com}
    & \tau_{\Pi} = 1/|\Pi|,
\end{align}
respectively. 

The distribution of $H_{ohm}$, $Q_{con}$, and $\Pi$ along the vertical axis at $x=0$ for $t=137.39$~s is shown in Figure \ref{fig:Temp-cause}, which also includes a temperature profile along the same cut. We notice that the temperature all along the midplane $x=0$ at this time is larger than the initial uniform temperature ($1$ MK), and has a maximum value of $\approx 3.9$ MK. This matches the findings of previous sections (e.g., Figure~\ref{fig:combined}, bottom panel in the fourth column, and Figure~\ref{fig:3dFL}, right panel). The merging of the two plasmoids leads to the formation of a small horizontal CS at $y\approx 26$ Mm (as discussed in section \ref{subsec:plasmoid}), and small localized enhanced current density structures (see the bottom panel in the second column of Figure \ref{fig:combined}). This high current region contributes to high Ohmic heating and corresponding temperature enhancement. This is reflected in Figure \ref{fig:Temp-cause}, where we see the near-spatial coincidence of a marked spike in the Ohmic heating and temperature curves at $y=26$~Mm. Magnetic field-aligned thermal conduction causes a redistribution of the thermal energy, which leads to the homogenization of the temperature along the field lines in the merged plasmoid. As a result, the enhanced temperature region appears to be a rim-like structure around the merging plasmoids with a horizontal strip crossing through the middle as shown in the top right panel of Figure~\ref{fig:3dFL} and in the middle panel of Figure \ref{fig:pi_Te_xy}. The temperature along the horizontal cut at $y=26$ Mm, which passes through the horizontal strip region, has a dip near $x=0$ (Figure~\ref{fig:pi_Te_xy}, blue curve in the right panel). The reason for the rim in the periphery of the plasmoid to be hotter than the horizontal patch region is that the rim is created during the 2D phase of reconnection, with temperatures at the rim of the upper plasmoid reaching is around $4$ MK at $t=81.58$ s, as shown in Figure \ref{fig:combined}, top panel in the fourth column. In the advanced merging state (at $t=137.39$ s) the thermal deposition due to conduction is more intense in the rim than in the horizontal strip region where the guide field ($B_z$) diminishes the effective temperature gradient along the field lines (see the field lines in Figure \ref{fig:3dFL}). Similarly, the absence of a dip in the temperature distribution at $x=0$ along the $y=27.5$ Mm (green) line in the right panel in Figure~\ref{fig:pi_Te_xy} is because the ellipse is nearly isothermal as a consequence of the thermal conduction. Another point worth noticing concerns the downward moving (primary) plasmoid; when it approaches the lower (primary) plasmoid, the plasma on the trailing side of the former has large $v_x$ (around $100$ km s$^{-1}$), and $v_y$ (around $300$ km s$^{-1}$) flows and expands (negative $\Pi$) as apparent in the left panel of Figure \ref{fig:pi_Te_xy}. This leads to a reduction in the temperature (see the dark patches at $y\approx 28.5$~Mm in the middle panel of Figure \ref{fig:pi_Te_xy}). The minimum temperature in those patches is $\approx 0.8$ MK (as shown by the red curve in the right panel of Figure \ref{fig:pi_Te_xy}), which is less than the initial isothermal temperature of 1 MK. 

To estimate the role of Ohmic heating, thermal conduction, and expansion/contraction in the temperature evolution in the region with appreciable electric currents, we consider the characteristic time scales $\tau_{ohm}, \tau_{con}$ and $\tau_{\Pi}$ in it. For that, we first mark the pixels in which the absolute current density, $|J|$, goes above a threshold value of $J_{th}$, which is taken as $5\%$ of the peak value of $|J|$ at the initial state ($t=0$); we will call this area the {\it ``Threshold Exceeding Current Zone''} (TECZ) in the rest of this section. Then we calculate the fractional counts of pixels that satisfy the criteria $\tau_{ohm} > \tau_{\Pi}$,  $\tau_{ohm} > \tau_{con}$, or $\tau_{\Pi} > \tau_{con}$. The time evolution of these fractional counts is shown in Figure \ref{fig:timescales-evolution}. The blue curve shows the fractional count of pixels with $\tau_{ohm} > \tau_{\Pi}$. The initial values correspond to the velocity map imposed as initial condition; at that time, $16.6$\% of the region selected for this calculation evolves with a stronger influence by the contraction associated with the initial velocity pulse than because of Ohmic heating; this fraction falls sharply in the first $5$~seconds of the simulation as the effect of the velocity pulse passes away from the CS region, and remains within the range $\approx 1-5\%$ till $t\approx 60$ s. Thereafter, when the plasmoids are gradually developed, we see that the area with predominance of the expansion/contraction increases and gradually occupies the majority of the region up to $76.8\%$ at the end of the simulation. For the $\tau_{ohm} > \tau_{con}$, and $\tau_{\Pi} > \tau_{con}$ curves (in red and green, respectively), we see that conduction is a slower process than both Ohmic heating and expansion/contraction in the entire region at the very beginning of the evolution, as a consequence of the initial uniform-temperature condition. Heat conduction becomes effective when important temperature inhomogeneities appear in the medium; the area with predominance of conduction gradually increases, in the comparison with ohmic heating and expansion/compression. From around $t=100$ s onward, the majority (more than $50\%$) of the region selected for this comparison has shorter characteristic time for heat conduction than for Ohmic heating and expansion/compression. The corresponding fractional regions gradually increase up to $\approx 84.3\%$ and $76.8\%$ at the end of the simulation for the $\tau_{ohm}>\tau_{con}$ (red curve) and $\tau_{\Pi}>\tau_{con}$ (green curve) criteria, respectively. We can also compare the fractional regions for the predominance of the different timescales at two of the important stages in the calculation: the premerger state at $t=81.58$ s, and the advanced merging state at $t=137.39$ s, as marked by the two vertical dashed lines in Figure~\ref{fig:timescales-evolution}. For the $\tau_{\Pi}>\tau_{con}$ criteria, the value increases from $46\%$ to $70\%$ from the premerger to the advanced merging states. Similarly, for the other two criteria: the fraction increases from $32.2\%$ to $64.6\%$ (for $\tau_{ohm}>\tau_{\Pi}$), and from $34\%$ to $73.3\%$ (for $\tau_{ohm}>\tau_{con}$) from the premerger to advanced merging states.

To show the relative importance of the three processes that can cause a change in temperature in the different regions, we calculate the spatial distribution of the ratios $\tau_{\Pi}/\tau_{con}$, $\tau_{ohm}/\tau_{\Pi}$, and $\tau_{ohm}/\tau_{\Pi}$ for the pixels in the TECZ, i.e., those that satisfy the condition $|J| > J_{th}$. We carry out that calculation for two time instances: first, the pre-merger stage, at $t=81.58$ s, and then in the advanced merging state at $t=137.39$ s. The results are shown in Figure \ref{fig:timescales-compare}. For $\tau_{\Pi}/\tau_{con}$ (left column, top panel for $t=81.58$~s and bottom panel for $t=137.39$~s), we see that the most notable expansion/contraction dominated regions (deep blue areas) are confined within the plasmoids before the merging occurs (at $t=81.58$ s), and within the baby-plasmoid region (between $y\approx 24.5-27$ Mm around $x=0$) in the advanced merged state (at $t=137.39$ s). This is because the islands are magnetically isolated, and therefore conduction becomes less effective than expansion/contraction within it. A significant predominance of thermal conduction over expansion/compression is seen at the rim of the top and bottom plasmoids at the premerger state, and in some parts of the periphery region of the plasmoid at the advanced merging state. For the ratio $\tau_{ohm}/\tau_{\Pi}$ (middle column), the dominance of expansion / contraction compared to ohmic heating (regions with red hues in the map) is observed at the necks  of the plasmoids (around $x=0$, at $y \approx 25, 27.5, 30.5$, and 33.5 Mm) for the pre-merger state (top row, middle panel), and mostly within the baby islands, and periphery parts of the merged plasmoid when it expands laterally during the merging process, at $t=137.39$ s. For $\tau_{ohm}/\tau_{con}$ (right column), because of the magnetically isolated structure in the pre-merger state (top-right panel), the magnetic field-aligned conduction is less  effective than ohmic heating (blue colored areas) inside the top and bottom plasmoids; instead, the situation is reversed (red-colored areas) along the rim of the top and plasmoids. Gradually, in the advanced merging state ($t=137.39$ s), the dominance of conduction compared to ohmic heating  occurs in most of the rim regions of the merged plasmoid; yet, the interface between the upper and lower plasmoids near $x=0$ is ohmic-heating dominated as shown in the bottom right panel due to the presence of the horizontal CS there (see also the second panel in the bottom row of Figure \ref{fig:combined}). 

The actual evolutionary timescales for the temperature vary from less than one second to more than $1000$~seconds in the different regions of the domain. To facilitate the identification, we present in Figure~\ref{fig:timescales} color maps for the timescales themselves, i.e., $\tau_{\Pi}$, $\tau_{ohm}$, and $\tau_{con}$, in the \textit{TECZ} regions, again for the premerger and advanced merging states (at $t=81.58$ and $137.39$ s, respectively). For the pre-merger state (upper row) we clearly identify that the thermal evolution at the rim of the upper and lower plasmoids is determined by the heat conduction (curved segments in blue, shown in the top-right panel in Figure~\ref{fig:timescales-compare}), with a short timescale, below $1$~s. At that stage, we can also identify the interior of the two plasmoids as regions in which the contraction/expansion, and ohmic heating timescales are determining the thermal evolution, although, in this case, the associated timescale is comparatively long, of order $10^2$~s. In the advanced stage, we see the predominance of the conduction timescale (which is below $1$~s) at the rim of a plasmoid that has formed at around $y=37.5$ Mm, and also in some regions at the periphery of the merged plasmoid apparent through the blue regions in the bottom-right panel of Figure~\ref{fig:timescales}. 

\section{Discussion and summary}\label{sec:summary}

In this paper, highly-collimated flows have been identified coming out of the reconnection site at the interface between the merging plasmoids. In the pre-merger state at $t=120.22$ s, when the two primary plasmoids are colliding with each other and creating a horizontal interface between them, we see a reconnection-driven bi-directional outflow along the horizontal direction (see Fig. \ref{fig:2d-3d-reconnection-transition}, right panel, and Fig.~\ref{fig:reconnection_2D}). The maximum velocity of this flow ($v_x$ component) is $\approx 100$ km s$^{-1}$, which reduces to $\lessapprox 50$ km s$^{-1}$ some $520$ km upward and downward of the interface (not shown). This implies that the flow is collimated along the $x$ direction; it extends from $x\approx 0.75$ to $\pm 4.5$ Mm, where it becomes much smaller. At a later time, $t=137.39$ s, the outward velocity decreases from its maximum value of $\approx 100$ km s$^{-1}$ (at $x=\pm 0.75$~Mm) to zero at approximately
$x=\pm 1.5$ Mm (and remains significantly low, with a few km~s$^{-1}$, at larger distances). This indicates that these collimated outflows are
short-lived, lasting for $\approx 17$ s, during which the spatial extent of the flow reduces from about $3.75$ to $0.75$~Mm. However, this sharp reconnection-driven outflow remains there till $t \approx 145$ s, and then reduces to $\lessapprox 30$ km s$^{-1}$. The temperature of the outflowing plasma is in the range between $\approx 1-4$ MK. We can also calculate the ideal-MHD terms in the total energy flux, namely
\begin{align}
    {\bf F_E} = \bigg(\frac{1}{2} \rho v^2 + \epsilon + p\bigg) {\bf v} - \frac{1}{4\pi} ({\bf v} \times {\bf B})\times {\bf B},
\end{align} along the $+x-$direction in the outflow regions where it reaches its maximum. To estimate the energy transport rate, we multiply this flux by the area through which the energy is transported, assuming that the width of these flows along the invariant direction is the same as in the $x-y$ plane (approximately 520 km). This calculation is repeated from $t=120.22$ to $137.39$~s at intervals of $4.29$~s. Finally, we integrate this energy rate over this time span to estimate the total energy budget of these ejections. The result is $\approx 3.3 \times 10^{24}$ erg, which is in the nanoflare range. 

These ejections share various physical properties with the so-called \textit{nanojets}, which are small-scale, short-lived, collimated outbursts observed in the solar corona \citep{Patrick:2021, 2022:ramada-nanojet}. The explanation of the nanojet events proposed by the former authors is that they are ``a consequence of the slingshot effect from the magnetically tensed, curved magnetic field lines reconnecting at small angles.'' In contrast, in our MHD model the collimated and short-lived flows that can be   associated with the nanojets are a direct consequence of the reconnection taking place at the interface between coalescing plasmoids and occurs both in the 2D and 2.5D phases of the reconnection process. The observation of nanojets is challenging because of its demand of high spatial resolution and temporal cadence, and thus, there are very few observations reported to date \citep[e.g.,][]{Patrick:2021, 2022:ramada-nanojet}, which were carried out for coronal loop systems. The observations suggest that the plane-of-sky (POS) speed of the nanojets is between $100-200$~km~s$^{-1}$ \citep{Patrick:2021}, and $100-286$ km s$^{-1}$ \citep{2022:ramada-nanojet}, bi-directional, and perpendicular to the magnetic strands. The width and length of these jets are on the order of
500~km and $1-3.8$~Mm, respectively, and their lifetime is reported to be $\approx 15$~s or less than that. These values are in good agreement with our model prediction. Our present model thus motivates future observation campaigns to investigate the occurrence of nanojets in a coronal environment that contains sharp CSs.

Our present model is 2.5D and does not incorporate the effect of radiative loss and lower atmospheric coupling to the transition region and the chromosphere, and hence, we do not obtain any thermal runaway scenario in our simulation. Therefore, we do not see appearance of coronal rain in our simulation unlike in the nanojet observations reported in \cite{Patrick:2021}. Yet, there is observational evidence of nanojets in a blowout jet scenario with no associated coronal rain \citep{2022:ramada-nanojet}. The nanojet-like ejections in our simulation depend mainly on the magnetic reconnection due to the coalescence of flux ropes in a coronal CS. However, how the magnetic and thermodynamic behavior of those ejections change with the incorporation of radiative losses and different forms of background heating (steady and time-dependent) is an interesting aspect which can be explored in future in a more realistic 3D model. Exploration of different parameter regimes can also be interesting to investigate the possible initiation of tearing and thermal instability in a coupled fashion, which may modify the instability growth rate of the CS (as reported in \citealt{Sen2022}), and formation of (localized) cool condensations similar to coronal rain or filament/prominence in the vicinity of a tearing CS \citep{Sen2022, Sen2023, DeJonghe+Sen:2024}.

To investigate the effect of spatial resolution, we have also run the simulation by activating three additional levels of AMR, keeping all the other parameters same to achieve a higher (spatial) resolution such that the smallest cell size is $\approx 13$ km along each direction, and find a similar qualitative evolution of the system: i.e., formation of two primary flux ropes in the upper half of the CS, followed by their merging that leads to a reconnection-driven bi-directional nanojet-like ejection from the interface of the upper and the lower (primary) plasmoids. The current study contributes to the theoretical understanding of the plasmoid-mediated reconnection in coronal CS, the magnetic and thermodynamic evolution due to coalescence of two flux ropes (plasmoids), and a possible explanation of small-scale transients (nanojets), which are important aspects to resolve the coronal heating mystery.

We highlight the key results of this work in the following.
\begin{enumerate}
    \item We setup a 2.5D numerical MHD model of a CS embedded in a stratified atmosphere in the solar corona which includes a guide field, uniform resistivity, and thermal conduction. The initial magnetic and thermodynamic parameters selected for this simulation (see Section \ref{sec:setup}) are representative of a typical solar coronal medium. The equilibrium CS is perturbed by a velocity pulse at three different locations along the CS to initiate forced magnetic reconnection. This leads to the disintegration of the original CS, and triggers the tearing instability, with formation of multiple plasmoids as it evolves in the nonlinear regime. However, we mainly focus on the two primary plasmoids that form at the upper half of the CS, and the discussion in this work concerns mainly those two. 

    \item The flows associated with the tearing instability drag the guide field (i.e., the $B_z$ component) into the interior of the plasmoids. This leads to their conversion into flux ropes or solenoids with guide field pointing along the symmetry axis (i.e., along the $+z$ direction), surrounded by purely planar field. The solenoids then move along the CS with speeds in the range $\approx 34-290$ km s$^{-1}$, which is in good agreement with the observations of plasmoids in a coronal jet spire as reported by \cite{Joshi2020}, who estimated their speed to be between $10-220$ km s$^{-1}$. The temperature of the plasmoids in our model is between $0.8-4$ MK, which is also in fair agreement with the observations by \cite{Mulay2023}, who estimated their temperature to be $\approx 2.2$ MK in a jet spire region using the Differential Emission Measure (DEM) technique. In the experiment we find an additional, downward moving plasmoid that forms in the lower half of the CS (shown in Figure \ref{fig:combined}). This plasmoid is a consequence of reconnection-driven outflow, that carries the guide field trapped within it, and leaves a typical Harris CS (i.e., with nearly zero guide field) behind itself.

    \item When the plasmoids first meet, a secondary transverse CS is formed at their interface in which reconnection takes place. The transition from the initial stages ($t=120.22$ s) to the advanced merged stage ($t=139.37$ s) is accompanied by the change of character of the reconnection, namely, from simple 2D reconnection to 2.5D reconnection, i.e., 2D reconnection with a guide field. 

    \item During the merging phase of the two primary plasmoids, small, collimated bi-directional nanojet-like ejections are produced due to magnetic reconnection at the interface between the upper and lower plasmoids. These ejections share various characteristics  (size, velocity, duration, energy) with the nanojets observed by \cite{Patrick:2021, 2022:ramada-nanojet}.

    \item The temperature evolution within the CS is a combined effect of Ohmic heating, thermal conduction, and expansion/contraction of the plasma; which of them is predominant can be estimated from their respective characteristic time scales. The field-aligned thermal conduction distributes the thermal energy along the magnetic field lines. This leads to the appearance of a rim-like enhanced temperature region around the merging plasmoids at $t=137.39$ s with a horizontal patch at their interface (see middle panel of Figure \ref{fig:Temp-cause}). This temperature enhancement at the periphery of the merging plasmoids cannot be confirmed using any of the available observations to date, as it requires very high spatial resolution to resolve these structures. In future work we will produce synthetic maps to compare with measurements in the EUV in current satellite missions like the \textit{Atmospheric Imaging Assembly (AIA)} \citep{2012:sdo, 2012:aia}, or \textit{High Resolution Imager (HRI)} of \textit{Solar Orbiter} campaigns \citep{2020:rochus}.
\end{enumerate}      

\begin{acknowledgements}
The authors acknowledge support by the European Research Council through the Synergy Grant \#810218 (``The Whole Sun”, ERC-2018-SyG). They thankfully acknowledge the technical expertise and assistance provided by the Spanish Supercomputing Network (Red Espa\~{n}ola de Supercomputaci{\'o}n), as well as the computer resources used: the LaPalma Supercomputer, located at the Instituto de Astrof{\'i}sica de Canarias. Data visualization and analysis were performed using \href{https://visit-dav.github.io/visit-website/index.html}{Visit}, \href{https://yt-project.org/}{the yt-project}, and UCAR’s \href{https://www.vapor.ucar.edu/pages/vaporCitationPage.html}{VAPOR} \citep{VAPOR1, VAPOR2}. 
\end{acknowledgements}

 \bibliographystyle{aa} 
 \bibliography{ref}

\begin{thebibliography}{52}
\expandafter\ifx\csname natexlab\endcsname\relax\def\natexlab#1{#1}\fi

\bibitem[{{Antolin} {et~al.}(2021){Antolin}, {Pagano}, {Testa}, {Petralia}, \& {Reale}}]{Patrick:2021}
{Antolin}, P., {Pagano}, P., {Testa}, P., {Petralia}, A., \& {Reale}, F. 2021, Nature Astronomy, 5, 54

\bibitem[{Archontis \& Hood(2013)}]{Archontis_Hood_2013}
Archontis, V. \& Hood, A.~W. 2013, The Astrophysical Journal, 769, L21, publisher: IOP Publishing

\bibitem[{{B{\'a}rta} {et~al.}(2011){B{\'a}rta}, {B{\"u}chner}, {Karlick{\'y}}, \& {Sk{\'a}la}}]{Barta2011}
{B{\'a}rta}, M., {B{\"u}chner}, J., {Karlick{\'y}}, M., \& {Sk{\'a}la}, J. 2011, \apj, 737, 24

\bibitem[{{Bhattacharjee} {et~al.}(2009){Bhattacharjee}, {Huang}, {Yang}, \& {Rogers}}]{Bhattacharjee2009}
{Bhattacharjee}, A., {Huang}, Y.-M., {Yang}, H., \& {Rogers}, B. 2009, Physics of Plasmas, 16, 112102

\bibitem[{{De Jonghe} \& {Sen}(2024)}]{DeJonghe+Sen:2024}
{De Jonghe}, J. \& {Sen}, S. 2024, arXiv e-prints, arXiv:2412.07427

\bibitem[{{F{\'a}rn{\'\i}k} {et~al.}(1983){F{\'a}rn{\'\i}k}, {Kaastra}, {K{\'a}lm{\'a}n}, {Karlick{\'y}}, {Slottje}, \& {Valni{\v{c}}ek}}]{Farnik1983}
{F{\'a}rn{\'\i}k}, F., {Kaastra}, J., {K{\'a}lm{\'a}n}, B., {et~al.} 1983, \solphys, 89, 355

\bibitem[{{Furth} {et~al.}(1963){Furth}, {Killeen}, \& {Rosenbluth}}]{1963PhFl....6..459F}
{Furth}, H.~P., {Killeen}, J., \& {Rosenbluth}, M.~N. 1963, Physics of Fluids, 6, 459

\bibitem[{{Giovanelli}(1939)}]{1939ApJ....89..555G}
{Giovanelli}, R.~G. 1939, \apj, 89, 555

\bibitem[{{Giovanelli}(1947)}]{1947MNRAS.107..338G}
{Giovanelli}, R.~G. 1947, \mnras, 107, 338

\bibitem[{{Giovanelli}(1948)}]{1948MNRAS.108..163G}
{Giovanelli}, R.~G. 1948, \mnras, 108, 163

\bibitem[{{Gosling} {et~al.}(1995){Gosling}, {McComas}, {Phillips}, {Pizzo}, {Goldstein}, {Forsyth}, \& {Lepping}}]{1995GeoRL..22.1753G}
{Gosling}, J.~T., {McComas}, D.~J., {Phillips}, J.~L., {et~al.} 1995, \grl, 22, 1753

\bibitem[{{Hesse} \& {Cassak}(2020)}]{2020JGRA..12525935H}
{Hesse}, M. \& {Cassak}, P.~A. 2020, Journal of Geophysical Research (Space Physics), 125, e25935

\bibitem[{{Huang} \& {Bhattacharjee}(2010)}]{Huang2010}
{Huang}, Y.-M. \& {Bhattacharjee}, A. 2010, Physics of Plasmas, 17, 062104

\bibitem[{{Joshi} {et~al.}(2020){Joshi}, {Chandra}, {Schmieder}, {Moreno-Insertis}, {Aulanier}, {N{\'o}brega-Siverio}, \& {Devi}}]{Joshi2020}
{Joshi}, R., {Chandra}, R., {Schmieder}, B., {et~al.} 2020, \aap, 639, A22

\bibitem[{{Karpen} {et~al.}(2012){Karpen}, {Antiochos}, \& {DeVore}}]{2012ApJ...760...81K}
{Karpen}, J.~T., {Antiochos}, S.~K., \& {DeVore}, C.~R. 2012, \apj, 760, 81

\bibitem[{{Keppens} {et~al.}(2003){Keppens}, {Nool}, {T{\'o}th}, \& {Goedbloed}}]{Keppens:2003}
{Keppens}, R., {Nool}, M., {T{\'o}th}, G., \& {Goedbloed}, J.~P. 2003, Computer Physics Communications, 153, 317

\bibitem[{Keppens {et~al.}(2021)Keppens, Teunissen, Xia, \& Porth}]{keppens2021}
Keppens, R., Teunissen, J., Xia, C., \& Porth, O. 2021, Computers \& Mathematics with Applications, 81, 316, development and Application of Open-source Software for Problems with Numerical PDEs

\bibitem[{{Keppens, R.} {et~al.}(2023){Keppens, R.}, {Popescu Braileanu}, {Zhou}, {Ruan}, {Xia}, {Guo}, {Claes}, \& {Bacchini}}]{keppens2023}
{Keppens, R.}, {Popescu Braileanu}, B., {Zhou}, Y., {et~al.} 2023, A\&A, 673, A66

\bibitem[{{Lemen} {et~al.}(2012){Lemen}, {Title}, {Akin}, {Boerner}, {Chou}, {Drake}, {Duncan}, {Edwards}, {Friedlaender}, {Heyman}, {Hurlburt}, {Katz}, {Kushner}, {Levay}, {Lindgren}, {Mathur}, {McFeaters}, {Mitchell}, {Rehse}, {Schrijver}, {Springer}, {Stern}, {Tarbell}, {Wuelser}, {Wolfson}, {Yanari}, {Bookbinder}, {Cheimets}, {Caldwell}, {Deluca}, {Gates}, {Golub}, {Park}, {Podgorski}, {Bush}, {Scherrer}, {Gummin}, {Smith}, {Auker}, {Jerram}, {Pool}, {Soufli}, {Windt}, {Beardsley}, {Clapp}, {Lang}, \& {Waltham}}]{2012:aia}
{Lemen}, J.~R., {Title}, A.~M., {Akin}, D.~J., {et~al.} 2012, \solphys, 275, 17

\bibitem[{Li {et~al.}(2019)Li, Jaroszynski, Pearse, Orf, \& Clyne}]{VAPOR1}
Li, S., Jaroszynski, S., Pearse, S., Orf, L., \& Clyne, J. 2019, Atmosphere, 10, 488

\bibitem[{{Lohner}(1987)}]{1987:Lohner}
{Lohner}, R. 1987, Computer Methods in Applied Mechanics and Engineering, 61, 323

\bibitem[{{Loureiro} {et~al.}(2007){Loureiro}, {Schekochihin}, \& {Cowley}}]{2007PhPl...14j0703L}
{Loureiro}, N.~F., {Schekochihin}, A.~A., \& {Cowley}, S.~C. 2007, Physics of Plasmas, 14, 100703

\bibitem[{{Mei} {et~al.}(2012){Mei}, {Shen}, {Wu}, {Lin}, {Murphy}, \& {Roussev}}]{Mei2012}
{Mei}, Z., {Shen}, C., {Wu}, N., {et~al.} 2012, \mnras, 425, 2824

\bibitem[{{Mondal} {et~al.}(2024){Mondal}, {Srivastava}, {Pontin}, {Yuan}, \& {Priest}}]{Mondal2024}
{Mondal}, S., {Srivastava}, A.~K., {Pontin}, D.~I., {Yuan}, D., \& {Priest}, E.~R. 2024, \apj, 963, 139

\bibitem[{{Moreno-Insertis} \& {Galsgaard}(2013)}]{Moreno-Insertis_Galsgaard:2013}
{Moreno-Insertis}, F. \& {Galsgaard}, K. 2013, \apj, 771, 20

\bibitem[{{Mulay} {et~al.}(2023){Mulay}, {Tripathi}, {Mason}, {Del Zanna}, \& {Archontis}}]{Mulay2023}
{Mulay}, S.~M., {Tripathi}, D., {Mason}, H., {Del Zanna}, G., \& {Archontis}, V. 2023, \mnras, 518, 2287

\bibitem[{{Ni} {et~al.}(2012){Ni}, {Roussev}, {Lin}, \& {Ziegler}}]{Ni2012}
{Ni}, L., {Roussev}, I.~I., {Lin}, J., \& {Ziegler}, U. 2012, \apj, 758, 20

\bibitem[{{Odstrcil} \& {Karlicky}(1997)}]{Odstrcil1997}
{Odstrcil}, D. \& {Karlicky}, M. 1997, \aap, 326, 1252

\bibitem[{{Ofman} \& {Liu}(2018)}]{Ofman2018}
{Ofman}, L. \& {Liu}, W. 2018, \apj, 860, 54

\bibitem[{{Pariat} {et~al.}(2009){Pariat}, {Antiochos}, \& {DeVore}}]{Pariat_etal:2009}
{Pariat}, E., {Antiochos}, S.~K., \& {DeVore}, C.~R. 2009, \apj, 691, 61

\bibitem[{Pearse {et~al.}(2023)Pearse, Li, clyne, StasJ, CoreCode, Daves, Hallock, Eroglu, Poplawski, \& Lacroix}]{VAPOR2}
Pearse, S., Li, S., clyne, {et~al.} 2023, NCAR/VAPOR: Vapor 3.8.1

\bibitem[{{Pesnell} {et~al.}(2012){Pesnell}, {Thompson}, \& {Chamberlin}}]{2012:sdo}
{Pesnell}, W.~D., {Thompson}, B.~J., \& {Chamberlin}, P.~C. 2012, \solphys, 275, 3

\bibitem[{{Popescu Braileanu} \& {Keppens}(2023)}]{Popescu_Keppens_2023}
{Popescu Braileanu}, B. \& {Keppens}, R. 2023, \aap, 678, A66

\bibitem[{{Porth} {et~al.}(2014){Porth}, {Xia}, {Hendrix}, {Moschou}, \& {Keppens}}]{2014ApJS..214....4P}
{Porth}, O., {Xia}, C., {Hendrix}, T., {Moschou}, S.~P., \& {Keppens}, R. 2014, \apjs, 214, 4

\bibitem[{{Potter} {et~al.}(2019){Potter}, {Browning}, \& {Gordovskyy}}]{Potter2019}
{Potter}, M.~A., {Browning}, P.~K., \& {Gordovskyy}, M. 2019, \aap, 623, A15

\bibitem[{{Priest} \& {Forbes}(2000)}]{2000mare.book.....P}
{Priest}, E. \& {Forbes}, T. 2000, {Magnetic Reconnection}

\bibitem[{{Rochus} {et~al.}(2020){Rochus}, {Auch{\`e}re}, {Berghmans}, {Harra}, {Schmutz}, {Sch{\"u}hle}, {Addison}, {Appourchaux}, {Aznar Cuadrado}, {Baker}, {Barbay}, {Bates}, {BenMoussa}, {Bergmann}, {Beurthe}, {Borgo}, {Bonte}, {Bouzit}, {Bradley}, {B{\"u}chel}, {Buchlin}, {B{\"u}chner}, {Cab{\'e}}, {Cadiergues}, {Chaigneau}, {Chares}, {Choque Cortez}, {Coker}, {Condamin}, {Coumar}, {Curdt}, {Cutler}, {Davies}, {Davison}, {Defise}, {Del Zanna}, {Delmotte}, {Delouille}, {Dolla}, {Dumesnil}, {D{\"u}rig}, {Enge}, {Fran{\c{c}}ois}, {Fourmond}, {Gillis}, {Giordanengo}, {Gissot}, {Green}, {Guerreiro}, {Guilbaud}, {Gyo}, {Haberreiter}, {Hafiz}, {Hailey}, {Halain}, {Hansotte}, {Hecquet}, {Heerlein}, {Hellin}, {Hemsley}, {Hermans}, {Hervier}, {Hochedez}, {Houbrechts}, {Ihsan}, {Jacques}, {J{\'e}r{\^o}me}, {Jones}, {Kahle}, {Kennedy}, {Klaproth}, {Kolleck}, {Koller}, {Kotsialos}, {Kraaikamp}, {Langer}, {Lawrenson}, {Le Clech'}, {Lenaerts}, {Liebecq}, {Linder}, {Long}, {Mampaey}, {Markiewicz-Innes}, {Marquet},
  {Marsch}, {Matthews}, {Mazy}, {Mazzoli}, {Meining}, {Meltchakov}, {Mercier}, {Meyer}, {Monecke}, {Monfort}, {Morinaud}, {Moron}, {Mountney}, {M{\"u}ller}, {Nicula}, {Parenti}, {Peter}, {Pfiffner}, {Philippon}, {Phillips}, {Plesseria}, {Pylyser}, {Rabecki}, {Ravet-Krill}, {Rebellato}, {Renotte}, {Rodriguez}, {Roose}, {Rosin}, {Rossi}, {Roth}, {Rouesnel}, {Roulliay}, {Rousseau}, {Ruane}, {Scanlan}, {Schlatter}, {Seaton}, {Silliman}, {Smit}, {Smith}, {Solanki}, {Spescha}, {Spencer}, {Stegen}, {Stockman}, {Szwec}, {Tamiatto}, {Tandy}, {Teriaca}, {Theobald}, {Tychon}, {van Driel-Gesztelyi}, {Verbeeck}, {Vial}, {Werner}, {West}, {Westwood}, {Wiegelmann}, {Willis}, {Winter}, {Zerr}, {Zhang}, \& {Zhukov}}]{2020:rochus}
{Rochus}, P., {Auch{\`e}re}, F., {Berghmans}, D., {et~al.} 2020, \aap, 642, A8

\bibitem[{{Sakai}(1983)}]{Sakai1983}
{Sakai}, J.~I. 1983, Journal of Plasma Physics, 30, 109

\bibitem[{{Sakai} \& {Washimi}(1982)}]{SakaiWashimi1982}
{Sakai}, J.~I. \& {Washimi}, H. 1982, \apj, 258, 823

\bibitem[{{Schl{\"u}ter}(1957)}]{Schluter_1957}
{Schl{\"u}ter}, A. 1957, in IAU Symposium, Vol.~4, Radio astronomy, ed. H.~C. {van de Hulst}, 356

\bibitem[{{Schmidt} \& {Cargill}(2003)}]{2003JGRA..108.1023S}
{Schmidt}, J.~M. \& {Cargill}, P.~J. 2003, Journal of Geophysical Research (Space Physics), 108, 1023

\bibitem[{{Schumacher} \& {Kliem}(1997)}]{Schumacher1997}
{Schumacher}, J. \& {Kliem}, B. 1997, Physics of Plasmas, 4, 3533

\bibitem[{{Sen} {et~al.}(2023){Sen}, {Jenkins}, \& {Keppens}}]{Sen2023}
{Sen}, S., {Jenkins}, J., \& {Keppens}, R. 2023, \aap, 678, A132

\bibitem[{{Sen} \& {Keppens}(2022)}]{Sen2022}
{Sen}, S. \& {Keppens}, R. 2022, \aap, 666, A28

\bibitem[{{Shen} {et~al.}(2011){Shen}, {Lin}, \& {Murphy}}]{Shen2011}
{Shen}, C., {Lin}, J., \& {Murphy}, N.~A. 2011, \apj, 737, 14

\bibitem[{{Sukarmadji} {et~al.}(2022){Sukarmadji}, {Antolin}, \& {McLaughlin}}]{2022:ramada-nanojet}
{Sukarmadji}, A. R.~C., {Antolin}, P., \& {McLaughlin}, J.~A. 2022, \apj, 934, 190

\bibitem[{{van Leer}(1974)}]{1974:vanLeer}
{van Leer}, B. 1974, Journal of Computational Physics, 14, 361

\bibitem[{{Wang} {et~al.}(2001){Wang}, {Chae}, {Yurchyshyn}, {Yang}, {Steinegger}, \& {Goode}}]{Wang2001}
{Wang}, H., {Chae}, J., {Yurchyshyn}, V., {et~al.} 2001, \apj, 559, 1171

\bibitem[{{Wyper} \& {DeVore}(2016)}]{Wyper_DeVore:2016}
{Wyper}, P.~F. \& {DeVore}, C.~R. 2016, \apj, 820, 77

\bibitem[{{Xia} {et~al.}(2018){Xia}, {Teunissen}, {El Mellah}, {Chan{\'e}}, \& {Keppens}}]{2018ApJS..234...30X}
{Xia}, C., {Teunissen}, J., {El Mellah}, I., {Chan{\'e}}, E., \& {Keppens}, R. 2018, \apjs, 234, 30

\bibitem[{{Yokoyama} \& {Shibata}(1996)}]{Yokoyama_Shibata:1996}
{Yokoyama}, T. \& {Shibata}, K. 1996, \pasj, 48, 353

\bibitem[{{Zhou} {et~al.}(2020){Zhou}, {Gao}, {Wang}, {Lin}, {Su}, {Jin}, \& {Zhang}}]{Zhou2020}
{Zhou}, G., {Gao}, G., {Wang}, J., {et~al.} 2020, \apj, 905, 150

\end{thebibliography}

\end{document}